\pdfoutput=1

\documentclass[pra,twocolumn,aps,amssymb,amsmath,tightenlines,floatfix]{revtex4}
\usepackage{graphicx,epsfig}

\newcommand{\cPT}{\ensuremath{\mathcal{PT}}}
\newcommand{\half}{\mbox{$\textstyle{\frac{1}{2}}$}}
\newcommand{\fourth}{\mbox{$\textstyle{\frac{1}{4}}$}}
\newcommand{\eighth}{\mbox{$\textstyle{\frac{1}{8}}$}}
\newcommand{\threehalf}{\mbox{$\textstyle{\frac{3}{2}}$}}
\newcommand{\fourthird}{\mbox{$\textstyle{\frac{4}{3}}$}}
\newcommand{\vep}{\varepsilon}

\begin{document}
\title{Behavior of eigenvalues in a region of broken-$\cPT$ symmetry}

\author{Carl M. Bender$^a$}\email{cmb@wustl.edu}
\author{Nima Hassanpour$^a$}\email{nimahassanpourghady@wustl.edu}
\author{Daniel W. Hook$^{a,b}$}\email{d.hook@digital-science.com}
\author{S.~P.~Klevansky$^c$}\email{spk@physik.uni-heidelberg.de}
\author{Christoph S\"underhauf$^{a,c}$}\email{Suenderhauf@stud.uni-heidelberg.de}
\author{Zichao Wen$^{a,d,e}$}\email{zcwen@amss.ac.cn}

\affiliation{$^a$Department of Physics, Washington University, St. Louis,
Missouri 63130, USA\\
$^b$Centre for Complexity Science, Imperial College London, London SW7 2AZ, UK\\
$^c$Institut f\"{u}r Theoretische Physik, Universit\"{a}t
Heidelberg, Philosophenweg 12, 69120 Heidelberg, Germany\\
$^d$University of Chinese Academy of Sciences, Beijing 100049, China,\\
$^e$Key Laboratory of Mathematics Mechanization, Institute of Systems Science,
AMSS, Chinese Academy of Sciences, Beijing 100190, China}

\begin{abstract}
$\cPT$-symmetric quantum mechanics began with a study of the
Hamiltonian $H=p^2+x^2(ix)^\vep$. When $\vep\geq0$, the eigenvalues of this
non-Hermitian Hamiltonian are discrete, real, and positive. This portion of
parameter space is known as the region of {\it unbroken} $\cPT$ symmetry. In the
region of {\it broken} $\cPT$ symmetry $\vep<0$ only a finite number of
eigenvalues are real and the remaining eigenvalues appear as complex-conjugate
pairs. The region of unbroken $\cPT$ symmetry has been studied but the region of
broken $\cPT$ symmetry has thus far been unexplored. This paper presents a
detailed numerical and analytical examination of the behavior of the eigenvalues
for $-4<\vep<0$. In particular, it reports the discovery of an infinite-order
exceptional point at $\vep=-1$, a transition from a discrete spectrum to a
partially continuous spectrum at $\vep=-2$, a transition at the Coulomb value
$\vep=-3$, and the behavior of the eigenvalues as $\vep$ approaches the
conformal limit $\vep=-4$.
\end{abstract}
\maketitle

\section{Introduction}
$\cPT$-symmetric quantum theory has its roots in a series of papers that
proposed a new perturbative approach to scalar quantum field theory: Instead of
a conventional expansion in powers of a coupling constant, it was proposed that
a perturbation parameter $\delta$ be introduced that measures the nonlinearity
of the theory. Thus, to solve a $g\phi^4$ field theory one studies a $g\phi^2(
\phi^2)^\delta$ theory and treats $\delta$ as a small parameter. After
developing a perturbation expansion in powers of $\delta$, the parameter
$\delta$ is set to one to obtain the results for the $g\phi^4$ theory. This
perturbative calculation is impressively accurate and does not require the
coupling constant $g$ to be small \cite{T1,T2}. A crucial technical feature of
this idea is that $\phi^2$ and not $\phi$ be raised to the power $\delta$ in
order to avoid raising a negative number to a noninteger power and thereby
generating complex numbers as an artifact of the procedure.

Subsequently, the $\delta$ expansion was used to solve an array of nonlinear
classical differential equations taken from various areas of physics: The
Thomas-Fermi equation (nuclear charge density) $y''(x)=[y(x)]^{3/2}/\sqrt{x}$ is
modified to $y''(x)=y(x)[y(x)/x]^\delta$; the Lane-Emdon equation (stellar
structure) $y''(x)+2y'(x)/x+[y(x)]^n=0$ is modified to $y''(x)+2y'(x)+[y(x)]^{1+
\delta}$; the Blasius equation (fluid dynamics) $y'''(x)+y''(x)y(x)=0$ is
modified to $y'''(x)+y''(x)[y(x)]^\delta=0$; the Korteweg-de Vries equation
(nonlinear waves) $u_t+uu_x+u_{xxx}=0$ is modified to $u_t+u^\delta u_x+u_{xxx}
=0$. In each of these cases the quantity raised to the power delta is positive
and when $\delta=0$ the equation becomes linear. Just a few terms in the
$\delta$ expansion gives an accurate numerical result \cite{T3}.

\begin{figure}[t!]
\begin{center}
\includegraphics[scale=0.21]{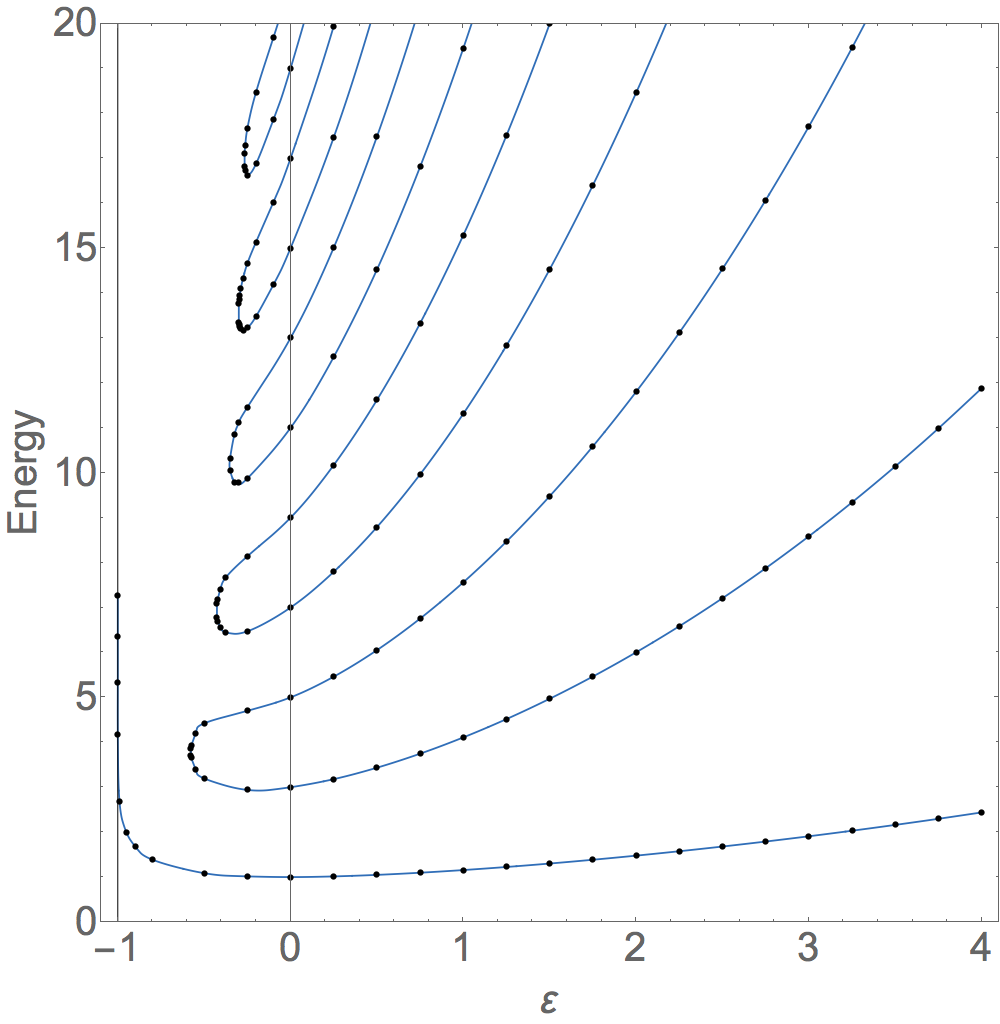}
\end{center}
\caption{[Color online] Real eigenvalues of the Hamiltonian $H=p^2+x^2(ix)^\vep$
plotted as functions of the parameter $\vep$. When $\vep\geq0$ (the region of
unbroken $\cPT$ symmetry), the spectrum is real, positive, and discrete.
However, as $\vep$ goes below $0$ ($\vep<0$ is known as the region of {\it
broken $\cPT$ symmetry}) the real eigenvalues begin to merge pairwise and form
complex-conjugate pairs. When $-1<\vep<0$, there are only a finite number of
real positive eigenvalues and an infinite number of complex-conjugate pairs of
eigenvalues. When $\vep\leq-0.57793$, only one real eigenvalue survives and as
$\vep$ approaches $-1^+$, this real eigenvalue becomes infinite. The behavior of
the complex eigenvalues in the region of broken $\cPT$ symmetry is not shown in
this graph and has not been explored until now.}
\label{F1}
\end{figure}

The breakthrough of $\cPT$-symmetric quantum theory was the surprising discovery
that to avoid the appearance of spurious complex numbers it is actually not
necessary to raise a positive quantity to the power $\delta$ so long as the
quantity is symmetric under combined space and time reflection. This fact is
highly nontrivial and was totally unexpected. For example, a quantum-mechanical
potential of form $x^2(ix)^\vep$ does not necessarily lead to complex
eigenvalues because the quantity $ix$ is $\cPT$ invariant. Indeed, the
non-Hermitian $\cPT$-symmetric Hamiltonian 
\begin{equation}
H=p^2+x^2(ix)^\vep
\label{e1}
\end{equation}
has the property that its eigenvalues are entirely real, positive, and discrete
when $\vep\geq0$ (see Fig.~\ref{F1}). The reality of the spectrum was noted in
Refs.~\cite{R1,R2} and was attributed to the $\cPT$ symmetry of $H$. Dorey,
Dunning, and Tateo proved that the  spectrum is real when $\vep>0$ \cite{R3,R4}.
Following the observation that the eigenvalues of non-Hermitian $\cPT$-symmetric
Hamiltonians could be real, many papers were published in which various
$\cPT$-symmetric model Hamiltonians were studied \cite{R5}.

A particularly interesting feature of $\cPT$-symmetric Hamiltonians is that they
often exhibit a transition from a parametric region of {\it unbroken} $\cPT$
symmetry in which all of the eigenvalues are real to a region of {\it broken}
$\cPT$ symmetry in which some of the eigenvalues are real and the rest of the
eigenvalues occur in complex-conjugate pairs. The $\cPT$ transition occurs in
both the classical and the quantized versions of a $\cPT$-symmetric Hamiltonian
\cite{R2} and this transition has been observed in numerous laboratory
experiments \cite{R6,R7,R8,R9,R10,R11,R12,R13,R14,R15,R16,R17}.

There have been many studies of the real spectrum of $H$ in (\ref{e1}) but
essentially nothing has been published regarding the analytic behavior of the
complex eigenvalues as functions of $\vep$ in the region of broken $\cPT$
symmetry. However, it is known that there is a sequence of negative-real values
of $\vep$ lying between $-1$ and $0$ at which pairs of real eigenvalues become
degenerate and split into pairs of complex-conjugate eigenvalues. These special
values of $\vep$ are often called {\it exceptional points} \cite{R18}. In
general, eigenvalues usually have square-root branch-point singularities at
exceptional points.

Exceptional points in the complex plane, sometimes called {\it Bender-Wu
singularities}, explain the divergence of perturbation expansions 
\cite{R19,R20}. The appearance of exceptional points is a generic phenomenon. In
these early studies of coupling-constant analyticity it was shown that the
energy levels of a Hamiltonian, such as the Hamiltonian for the quantum
anharmonic oscillator $H=p^2+x^2+gx^4$, are analytic continuations of one
another as functions of the complex coupling constant $g$ due to the phenomenon
of level crossing at the exceptional points. Thus, the energy levels of a
quantum system, which are discrete when $g$ is real and positive, are actually
smooth analytic continuations of one another in the complex-$g$ plane
\cite{R21}. A simple topological picture of quantization emerges: The discrete
energy levels of a Hamiltonian for $g>0$ are all branches of a multivalued
energy function $E(g)$ and the distinct eigenvalues of this Hamiltonian
correspond with the sheets of the Riemann surface on which $E(g)$ is defined.
Interestingly, it is possible to vary the parameters of a Hamiltonian in
laboratory experiments and thus to observe {\it experimentally} the effect of
encircling exceptional points \cite{R13,R22,R23}.

The purpose of this paper is to study the analytic continuation of the real
eigenvalues shown in Fig.~\ref{F1} as $\vep$ moves down the negative-$\vep$
axis. In Sec.~\ref{s2} we show that there is an {\it infinite-order} exceptional
point at $\vep=-1$ where there is an elaborate logarithmic spiral (a double
helix) of eigenvalues. The real part of each complex-conjugate pair of
eigenvalues that is formed at exceptional points between $\vep=-1$ and $\vep=0$
approaches $+\infty$ like $|\ln(\vep+1)|^{2/3}$ as $\vep$ approaches $-1$. In
contrast, the imaginary parts of each pair of eigenvalues vanish logarithmically
at $\vep=-1$. As $\vep$ goes below $-1$, the real parts of the eigenvalues once
again become finite and the imaginary parts of the eigenvalues rise up from 0.
As $\vep$ goes from just above to just below $-1$, the imaginary parts of the
eigenvalues appear to undergo discrete jumps but in fact they vary
continuously as functions of $\vep$.

In Sec.~\ref{s3} we discuss the Stokes wedges that characterize the eigenvalue
problem as $\vep$ goes below $-1$. We give plots of the eigenvalues in the
region $-2<\vep<-1$ and perform an asymptotic analysis of the eigenvalues near
$\vep=-2$. As $\vep$ approaches $-2$, the entire spectrum becomes degenerate;
the real parts of all the eigenvalues approach $-1$ and the imaginary parts
coalesce to $0$.

Section~\ref{s4} presents a numerical study of the eigenvalues in the region
$-4<\vep<-2$. We show that a transition occurs at $\epsilon=-2$ in
which the eigenspectrum goes from being discrete to becoming partially discrete
and partially {\it continuous}. The continuous part of the spectrum lies on
complex-conjugate pairs of curves in the complex-$\vep$ plane. Another
transition occurs at $\vep=-3$ (the $\cPT$-symmetric Coulomb potential); below
$\vep=-3$ some of the discrete eigenvalues become real. As $\vep$ approaches the
conformal point $\vep=-4$, the eigenvalues collapse to the single value 0.
Section~\ref{s5} gives brief concluding remarks.

\section{Eigenvalue behavior as $\vep\to-1$}
\label{s2}

\subsection{Stokes wedges}
The time-independent Schr\"odinger eigenvalue problem for the Hamiltonian $H$ in
(\ref{e1}) is characterized by the differential equation
\begin{equation}
-y''(x)+x^2(ix)^\vep y(x)=Ey(x).
\label{e2}
\end{equation}
The boundary conditions imposed on the eigenfunctions require that $y(x)\to0$
exponentially rapidly as $|x|\to\infty$ in a pair of Stokes wedges in the
complex-$x$ plane. This subsection explains the locations of these Stokes
wedges.

As has been previously discussed at length, the potential $x^2(ix)^\vep$ has a
logarithmic singularity in the complex-$x$ plane when $\vep$ is not an integer.
Thus, it is necessary to introduce a branch cut. This branch cut is chosen to
run from $0$ to $\infty$ in the complex-$x$ plane along the positive-imaginary
axis because this choice respects the $\cPT$ symmetry of the Hamiltonian. This
is because $\cPT$ symmetry translates into left-right symmetry in the
complex-$x$ plane (that is, mirror symmetry with respect to the imaginary-$x$
axis) \cite{R1,R2}. The argument of $x$ on the principal sheet (sheet 0 of the
Riemann surface) runs from $-3\pi/2$ to $\pi/2$. On sheet 1, $\pi/2<{\rm arg}\,
x<5\pi/2$, on sheet $-1$, $-7\pi/2<{\rm arg}\,x<-3\pi/2$, and so on.

As explained in Refs.~\cite{R1,R2}, the Stokes wedges in which the boundary
conditions on $y(x)$ are imposed are located in the complex-$x$ plane in a
$\cPT$-symmetric fashion. If $\vep=0$, the Stokes wedges have angular opening
$\pi/2$ and are centered about the positive-$x$ and negative $x$-axes on the
principal sheet of the Riemann surface. As $\vep$ increases from $0$, the wedges
get narrower and rotate downwards; as $\vep$ decreases from $0$, the Stokes
wedges get wider and rotate upwards. WKB analysis provides precise formulas for
the location of the center line of the Stokes wedges,
\begin{eqnarray}
\theta_{\rm right\,wedge,\,center}&=&-\frac{\vep}{8+2\vep}\pi,\nonumber\\
\theta_{\rm left\,wedge,\,center}&=&-\pi+\frac{\vep}{8+2\vep}\pi,
\label{e3}
\end{eqnarray}
the upper edges of the Stokes wedges,
\begin{eqnarray}
\theta_{\rm
right\,wedge,\,upper\,edge}&=&\frac{2-\vep}{8+2\vep}\pi,\nonumber\\
\theta_{\rm left\,wedge,\,upper\,edge}&=&-\pi-\frac{2-\vep}{8+2\vep}\pi,
\label{e4}
\end{eqnarray}
and the lower edges of the Stokes wedges,
\begin{eqnarray}
\theta_{\rm
right\,wedge,\,lower~edge}&=&-\frac{2+\vep}{8+2\vep}\pi,\nonumber\\
\theta_{\rm left\,wedge,\,lower\,edge}&=&-\pi+\frac{2+\vep}{8+2\vep}\pi.
\label{e5}
\end{eqnarray}

\begin{figure}[h!]
\begin{center}
\includegraphics[scale=.36]{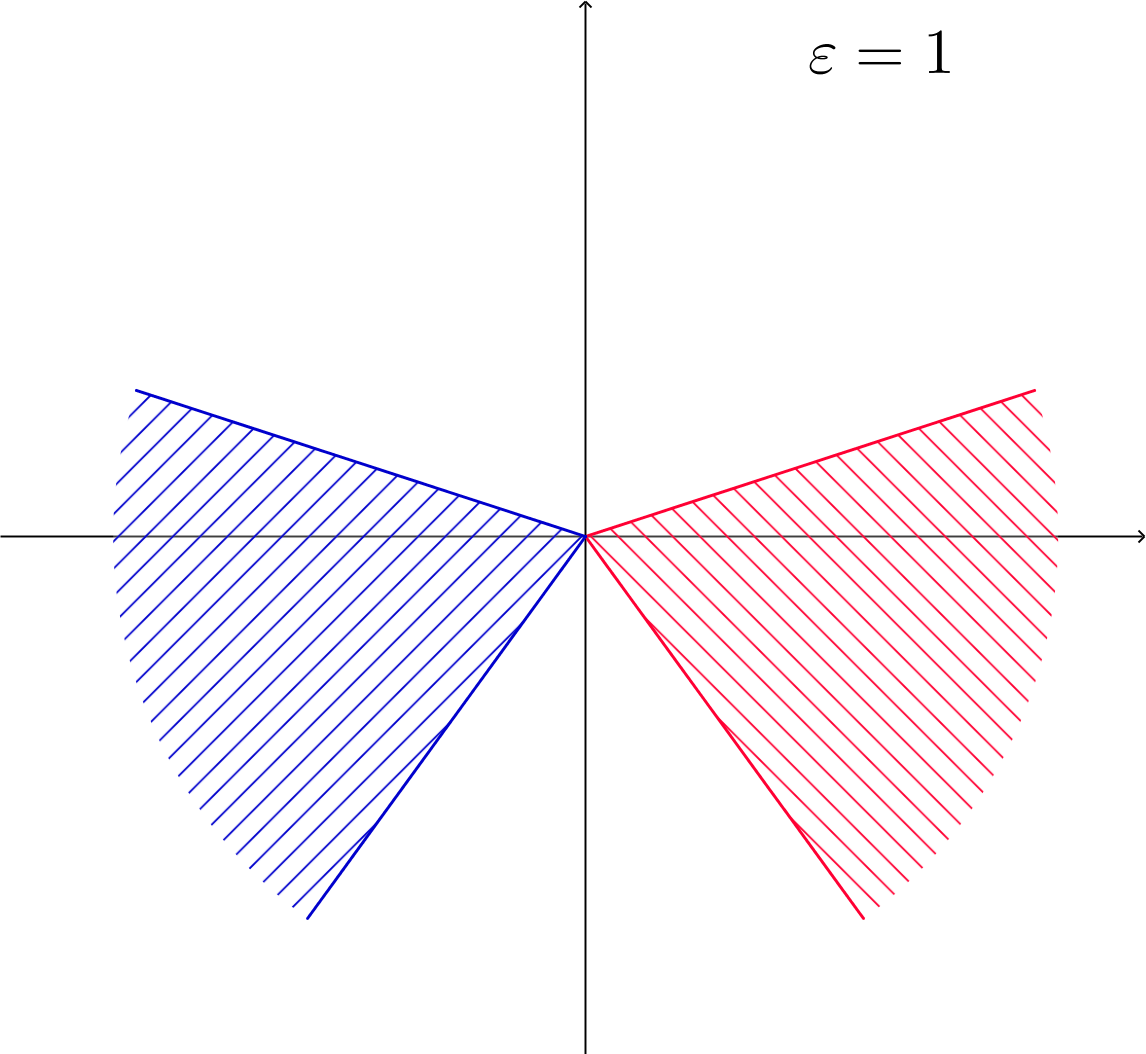}
\vspace{3mm}\includegraphics[scale=.36]{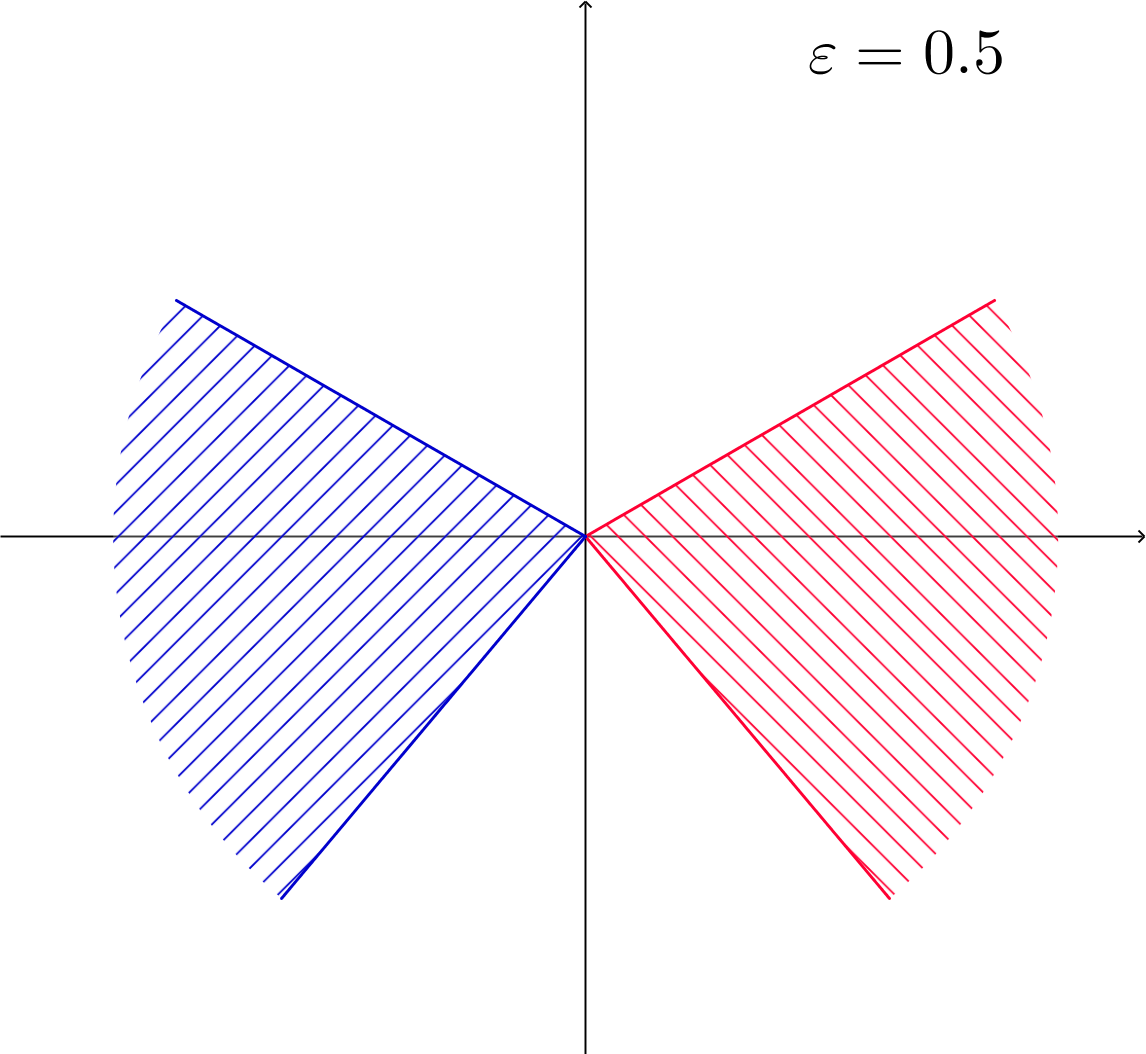}
\vspace{3mm}\includegraphics[scale=.36]{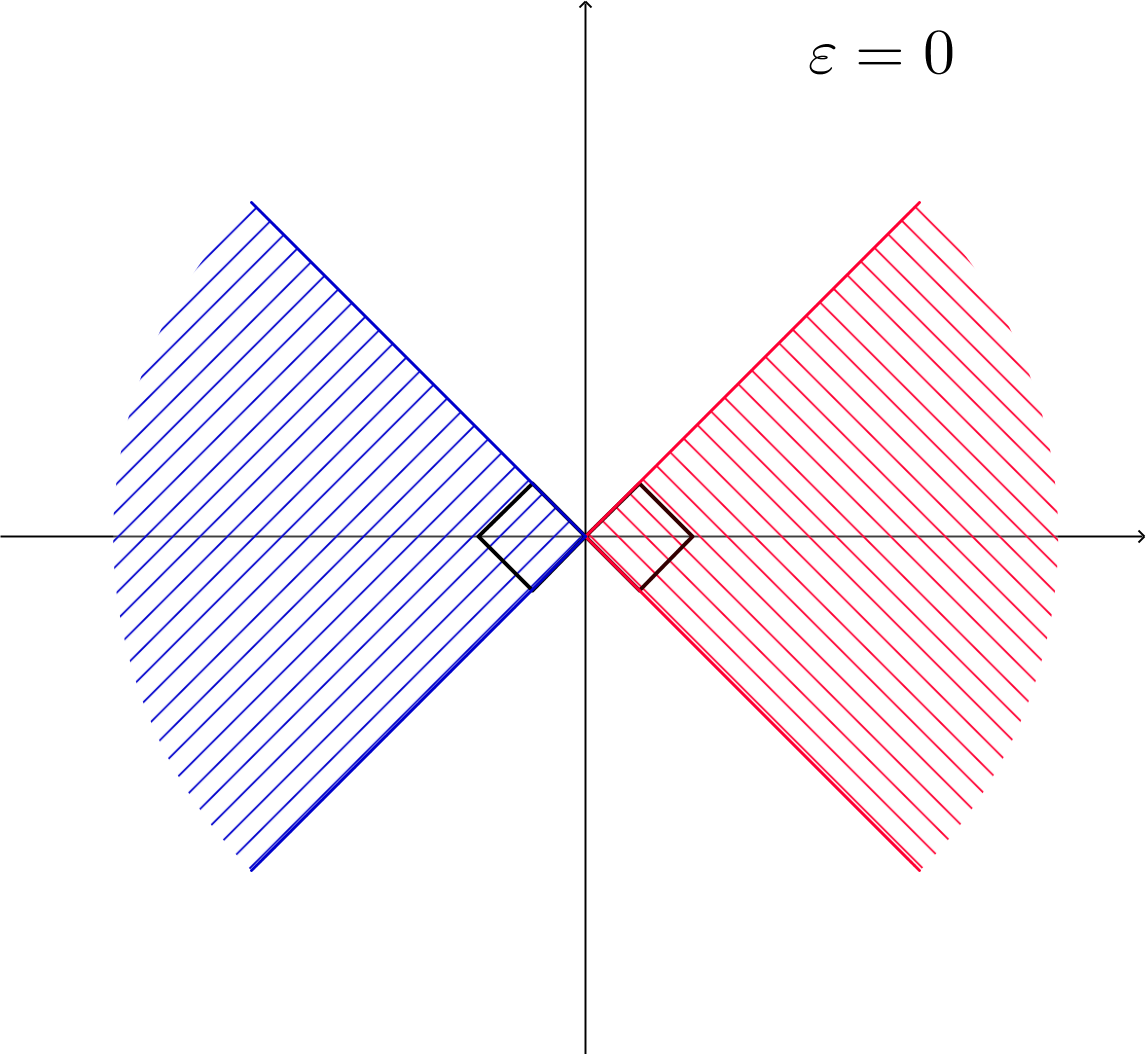}
\vspace{3mm}\includegraphics[scale=.36]{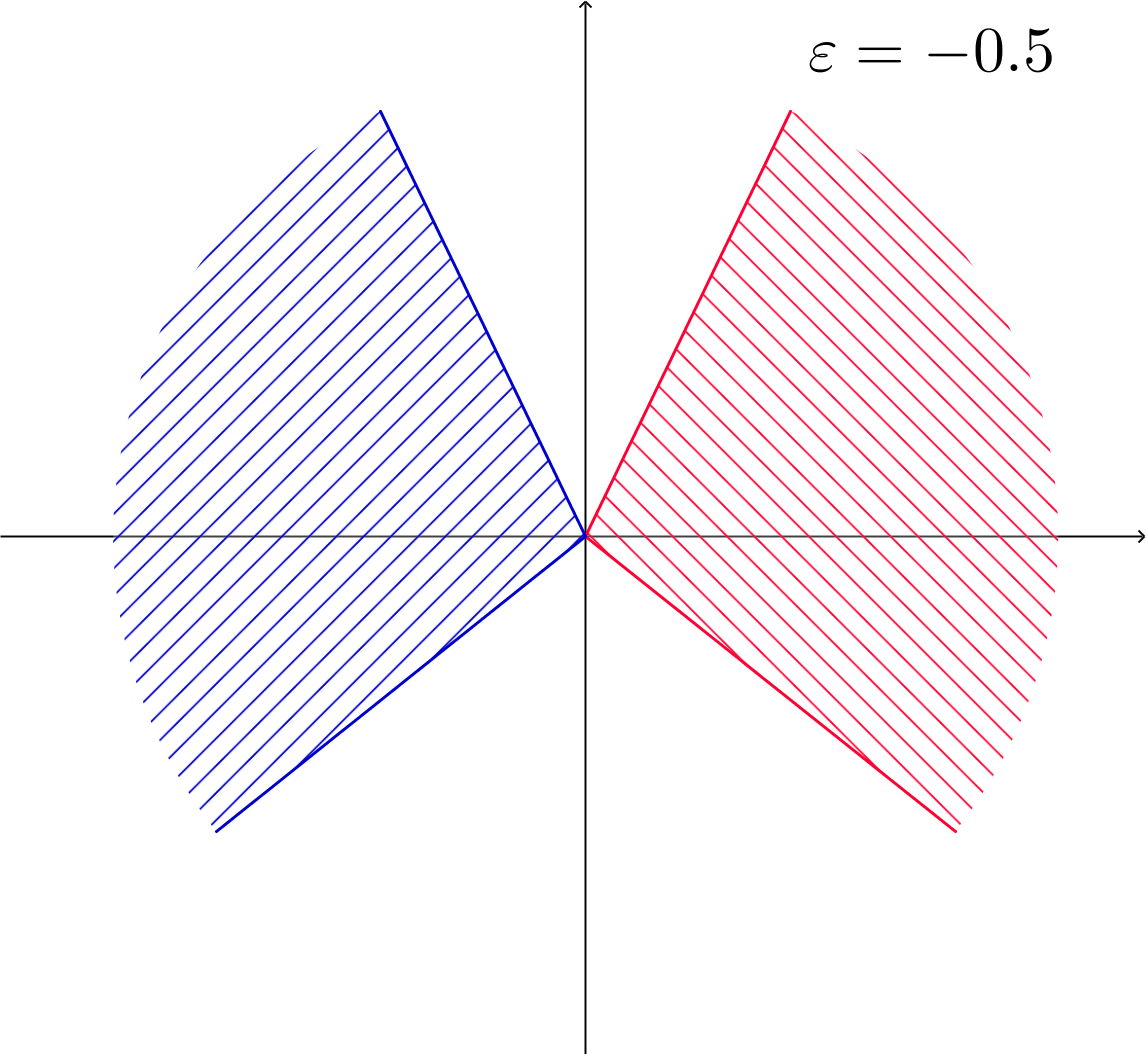}
\hspace{-4.5cm}
\end{center}
\begin{center}
\includegraphics[scale=.36]{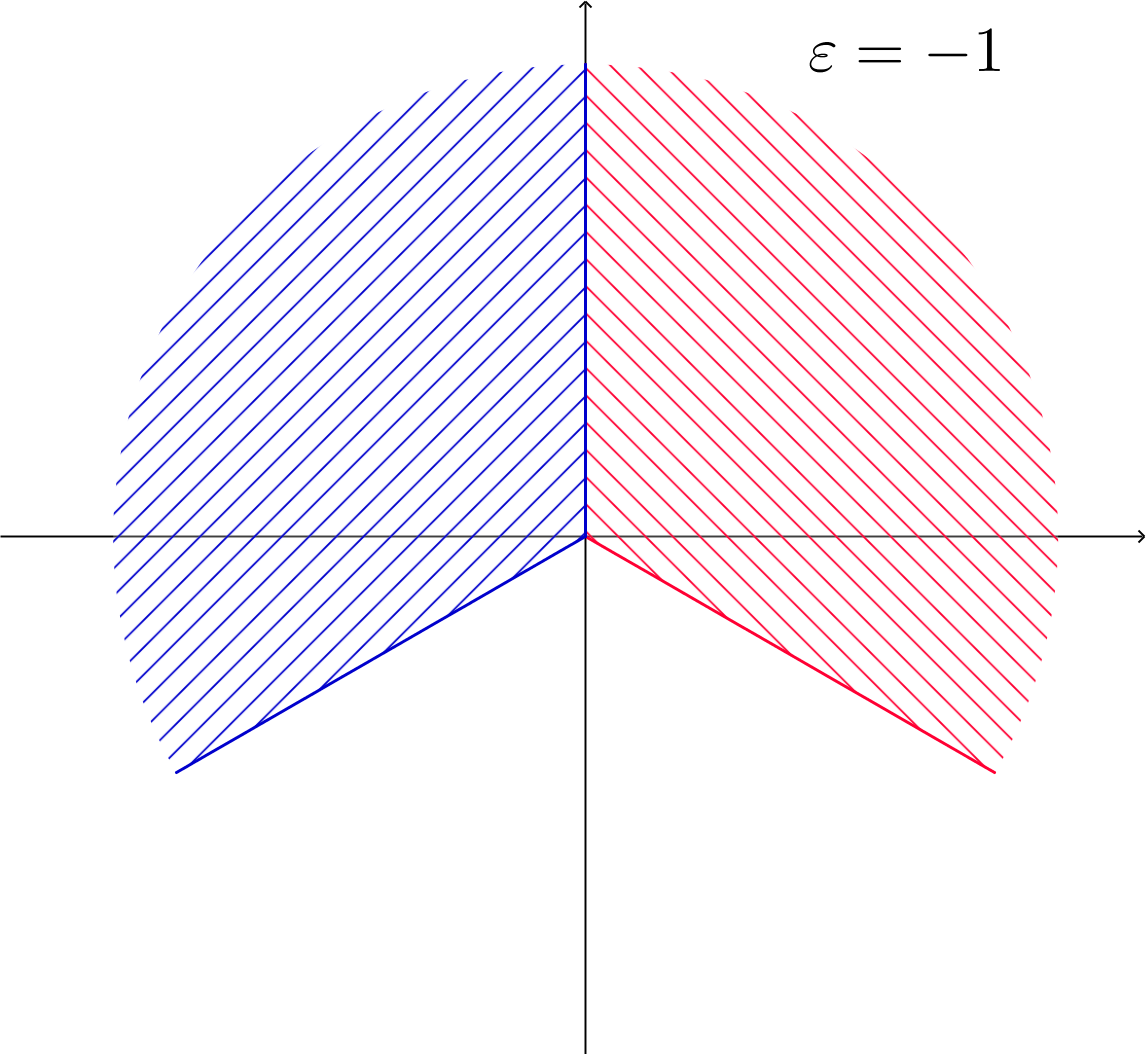}
\vspace{3mm}\includegraphics[scale=.36]{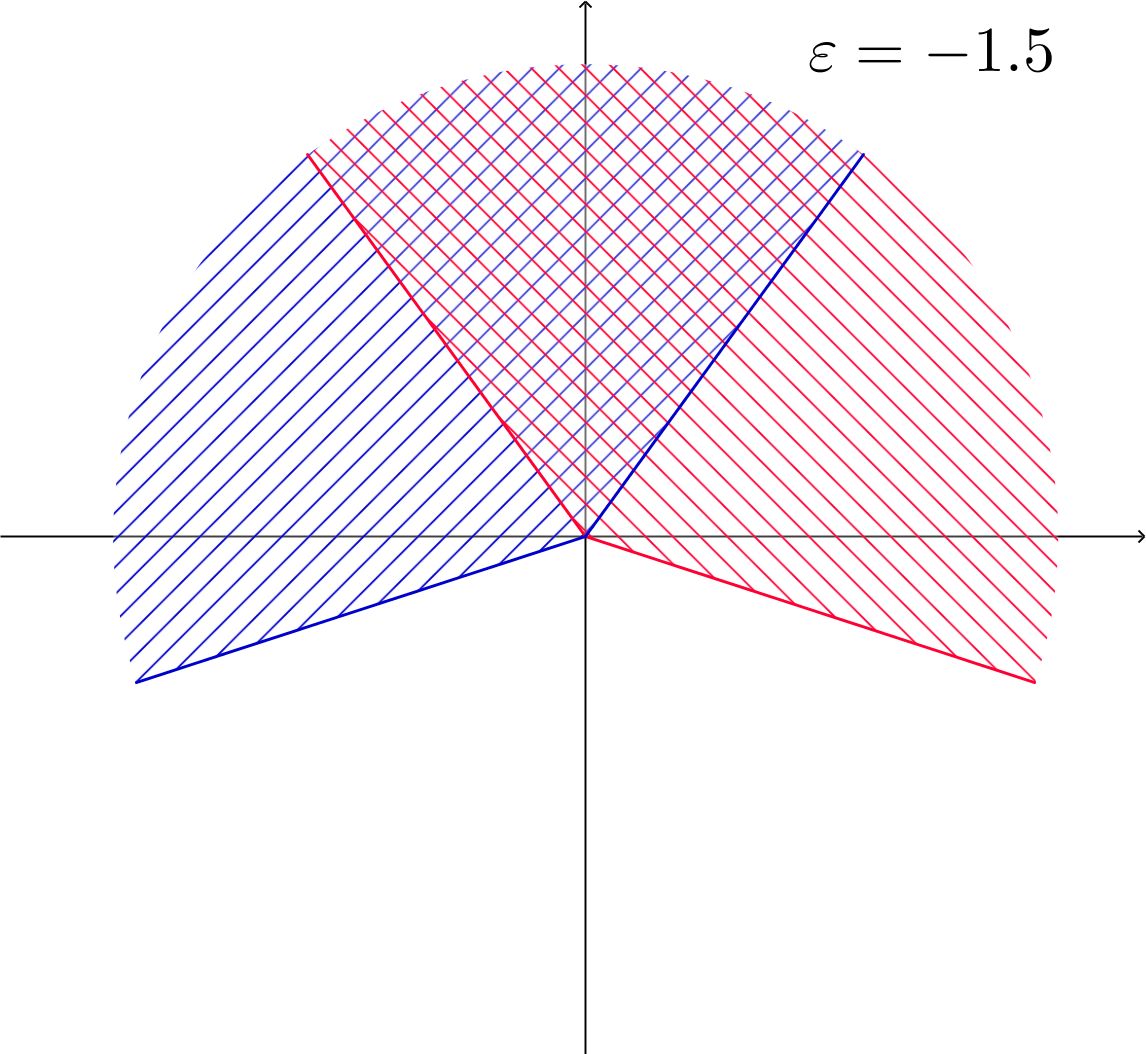}
\vspace{3mm}\includegraphics[scale=.36]{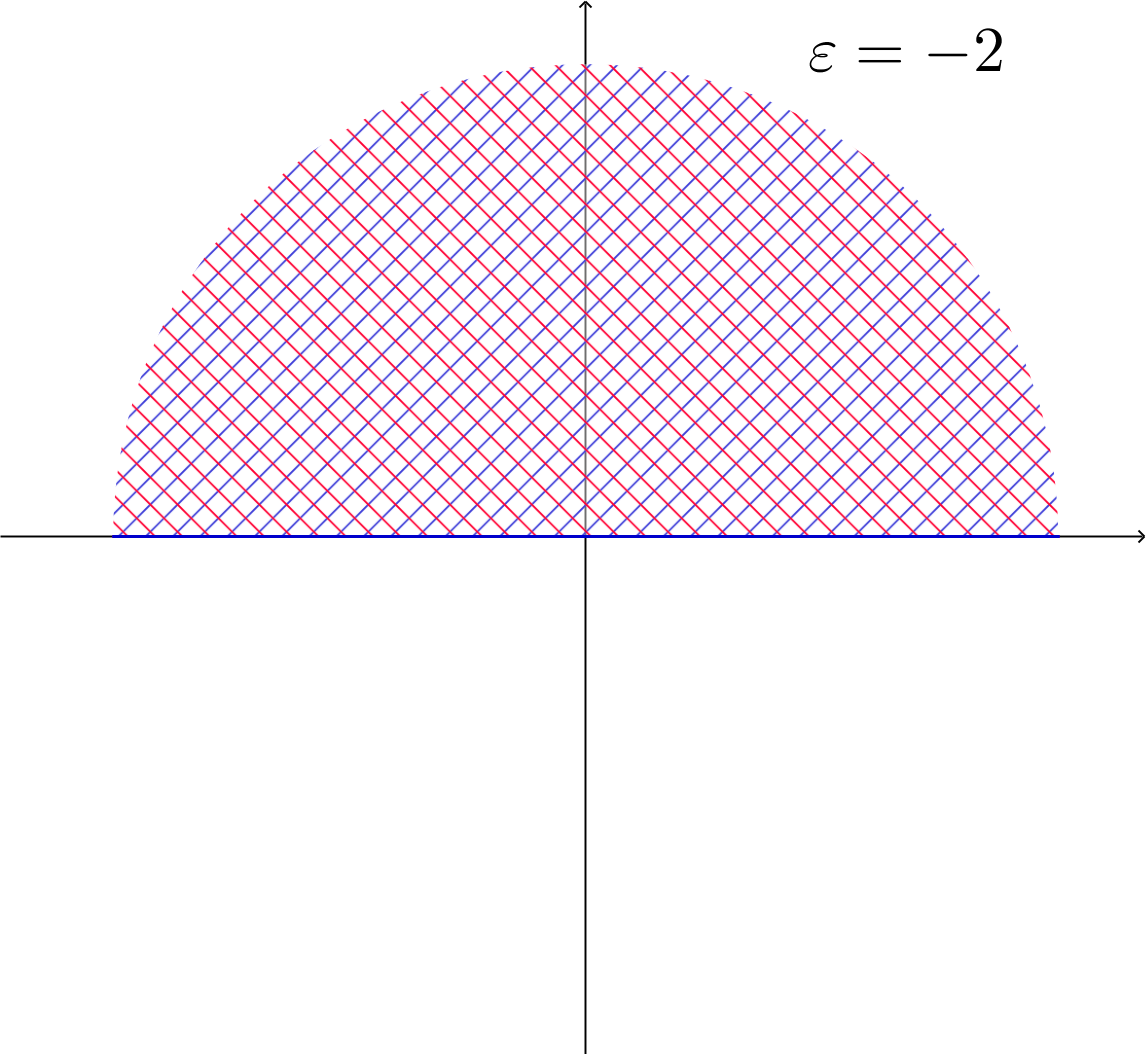}
\vspace{3mm}\includegraphics[scale=.36]{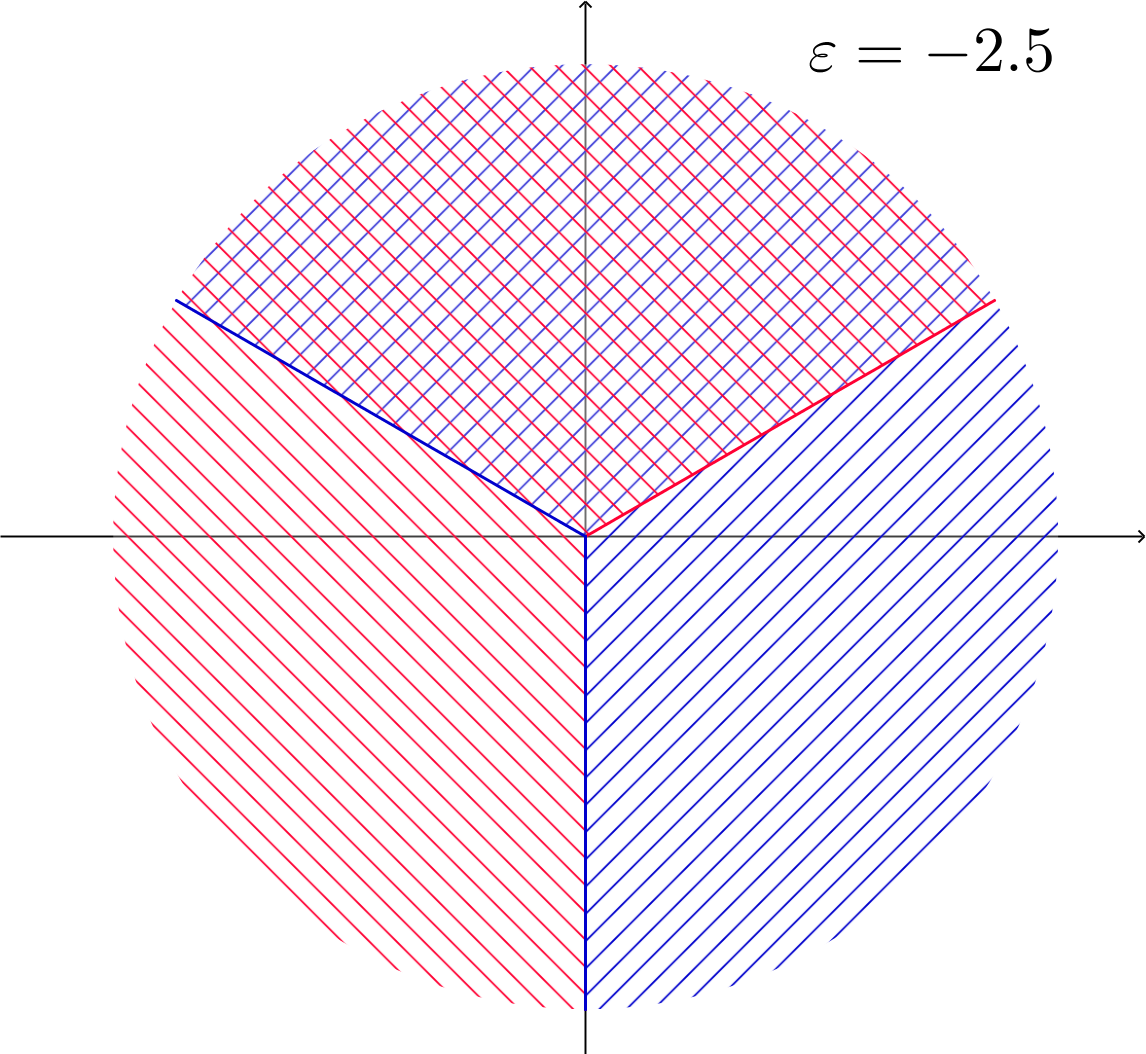}
% trim=left bottom right top
\end{center}
\caption{[Color online] Stokes wedges associated with the eigenvalue problem for
the Hamiltonian $H=p^2+x^2(ix)^\vep$ for eight values of $\vep$. The locations
of center lines, the upper edges, and the lower edges of the Stokes wedges are
given in (\ref{e3})-(\ref{e5}). The left wedge is colored blue and the right
wedge is colored red. As $\vep$ decreases, the wedges get wider and rotate
upwards. At $\vep=-1$ the two wedges touch and fuse into one wedge. However,
when $\vep$ goes below $-1$, the sheets are again distinct; the left wedge
rotates clockwise into sheet $-1$ and the right wedge rotates anticlockwise into
sheet $1$.}
\label{F2}
\end{figure}

The locations of the Stokes wedges for eight values of $\vep$ are shown in
Fig.~\ref{F2}. As $\vep$ decreases to $-1$, the opening angles of the wedges
increase to $120^\circ$ and the upper edges of the wedges touch. At the special
value $\vep=-1$ the logarithmic Riemann surface collapses to a single sheet; the
wedges fuse and are no longer separated. As a result there are no eigenvalues at
all (the spectrum is null) \cite{R25}. When $\vep$ goes below $-1$, the wedges
are again distinct and no longer touch; the left wedge rotates in the negative
direction and enters sheet $-1$ while the right wedge rotates in the positive
direction and enters sheet $1$.

\subsection{Numerical behavior of the eigenvalues as $\vep$ decreases below $0$}
Previous numerical studies of the (real) eigenvalues for $\vep\geq0$ were done
by using the shooting method. However, when the eigenvalues become complex, the
shooting method is not effective and we have used the finite-element method and
several variational methods. We have checked that the eigenvalues produced by
these different methods all agree to at least five decimal places.

\begin{figure}[t!]
\begin{center}
\includegraphics[trim=1mm 0mm 0mm 0mm,clip=true,scale=0.41]{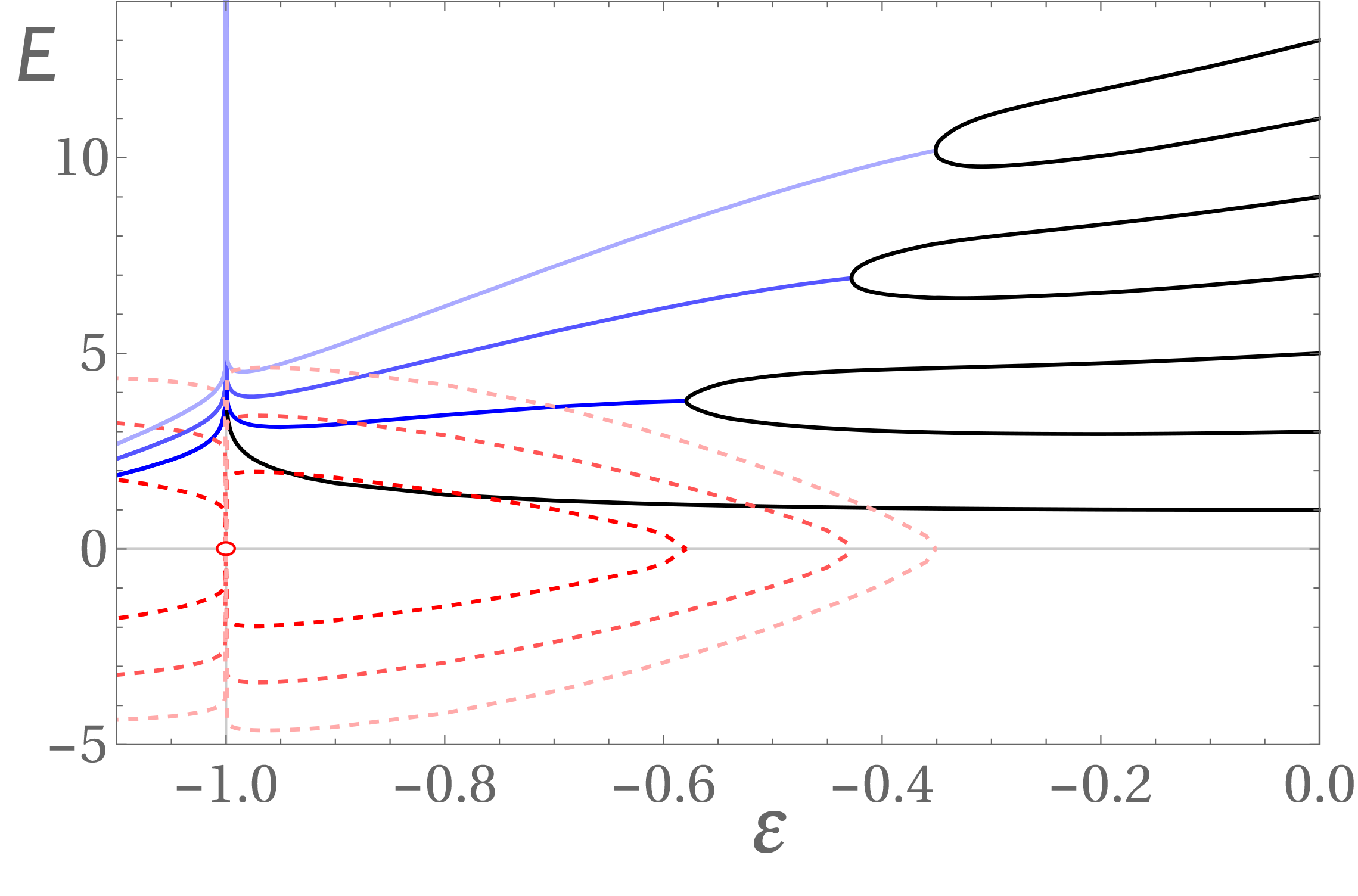}
% trim=left bottom right top
\end{center}
\caption{[Color online] Eigenvalues of the Hamiltonian $H=p^2+x^2(ix)^\vep$
plotted as functions of the parameter $\vep$ for $-1.1<\vep<0$. This graph is a
continuation of the graph in Fig.~\ref{F1}. As $\vep$ decreases below $0$
and enters the region of broken $\cPT$ symmetry, real eigenvalues (solid black
lines) become degenerate and then form complex-conjugate pairs. The real parts
of these pairs of eigenvalues (solid blue lines) initially decrease as $\vep$
decreases but blow up suddenly as $\vep$ approaches $-1$. The real parts then
decrease as $\vep$ decreases below $-1$. The imaginary parts of the eigenvalue
pairs (dashed red lines) remain finite and appear to suffer
discontinuous jumps at $\vep=-1$. However, a closer look shows that these dashed
lines rapidly decay to 0 near $\vep=-1$ and then rapidly come back up to
different values as $\vep$ passes through $-1$. A blow-up of the region near
$\vep=-1$ is given in Figs.~\ref{F4}.}
\label{F3}
\end{figure}

\begin{figure}[t!]
\begin{center}
\includegraphics[trim=1mm 0mm 0mm 0mm,clip=true,scale=0.395]{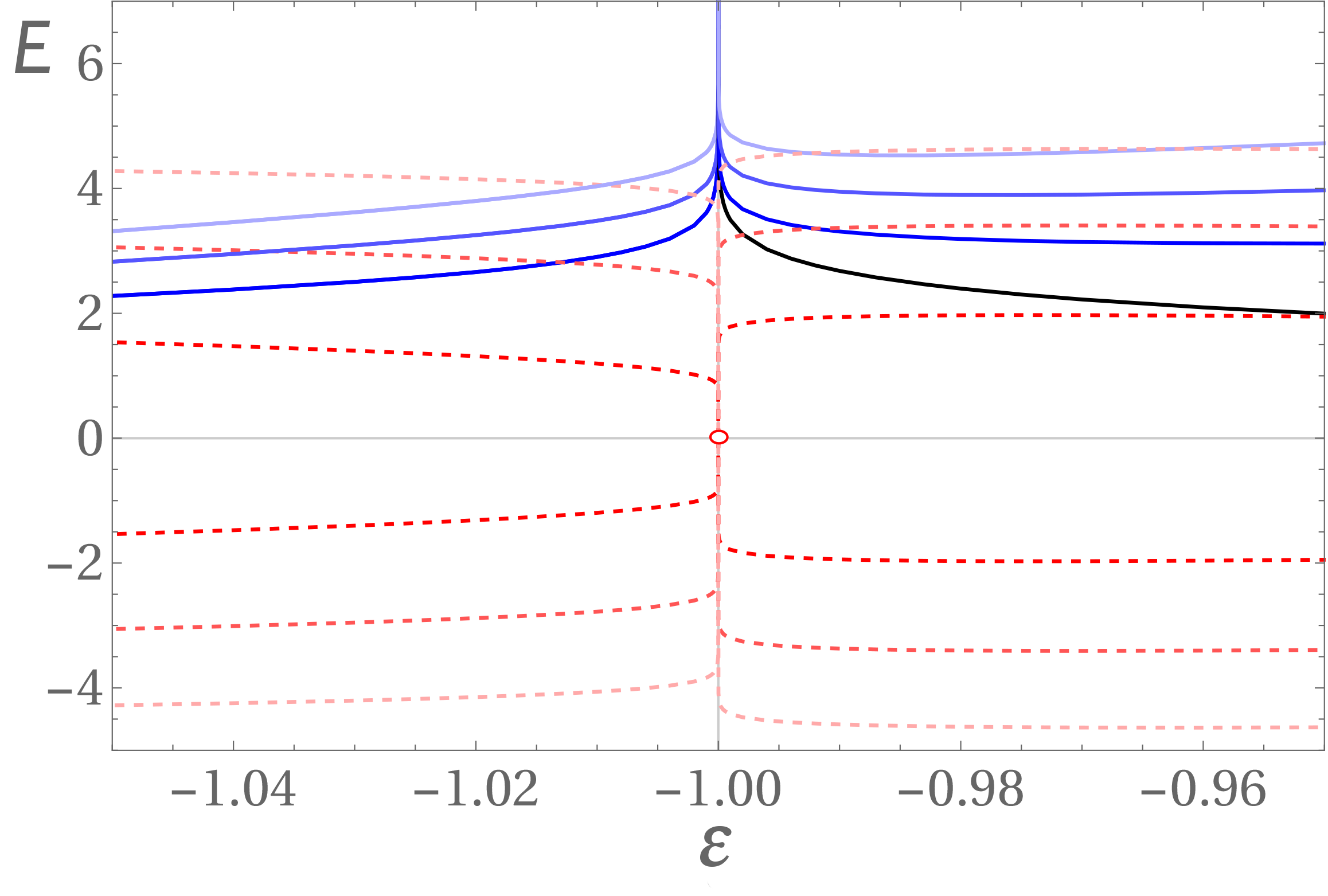}
% trim=left bottom right top
\end{center}
\caption{[Color online] Detailed view of Fig.~\ref{F3} showing the behavior of
the eigenvalues of the Hamiltonian $H=p^2+x^2(ix)^\vep$ plotted as functions of
the parameter $\vep$ for $-1.05\leq\vep\leq-0.95$. There is one real eigenvalue
for $\vep>-1$ (solid black line). The real parts of the complex eigenvalues
(blue solid lines) and the real eigenvalue diverge at $\vep=-1$. The complex
eigenvalues occur in complex-conjugate pairs and the imaginary parts of the
eigenvalues rapidly go to 0 at $\vep=-1$. These behaviors are expressed
quantitatively in (\ref{e14}).}
\label{F4}
\end{figure}

Figure \ref{F1} may seem to suggest that the real eigenvalues disappear pairwise
at special isolated values of $\vep$. However, the eigenvalues do not actually
disappear; rather, as each pair of real eigenvalues fuse, these eigenvalues
convert into a complex-conjugate pair of eigenvalues. At this transformation
point both the real and the imaginary parts of each pair of eigenvalues vary
{\it continuously}; the real parts remain nonzero and the imaginary parts move
away from zero as $\vep$ goes below the transition point. A more complete plot
of the eigenvalues in Fig.~\ref{F3} shows that the real parts of each pair of
eigenvalues decay slightly as $\vep$ decreases towards $-1$, while the imaginary
parts grow slowly in magnitude. However, just as $\vep$ reaches $-1$ the real
parts of the eigenvalues suddenly diverge logarithmically to $+\infty$ and the
imaginary parts of the eigenvalues suddenly vanish logarithmically. Below $\vep=
-1$ the real parts of the eigenvalues rapidly descend from $+\infty$ and the
imaginary parts of the eigenvalues rise up from $0$. This behavior is depicted
in Fig.~\ref{F3} and a detailed description of the region $-1.05<\vep<-0.95$ is
shown in Fig.~\ref{F4}.

\subsection{Asymptotic study of the eigenvalues near $\vep=-1$}
Figure \ref{F3} shows that the eigenvalues are singular at $\vep=-1$ and
suggests that this singularity is more complicated than the square-root
branch-point singularities that occur at standard exceptional points
\cite{R26}. To identify the singularity we perform a local
asymptotic analysis about the point $\vep=-1$. We begin by letting $\vep=-1+
\delta$ and we treat $\delta$ as small ($\delta<<1$). This allows us to
approximate the potential $x^2(ix)^\vep$ in (\ref{e1}) as
$$-ix\big[1+\delta\ln(ix)+{\rm O}\big(\delta^2 \big)\big].$$
We also expand the eigenfunctions in powers of $\delta$:
$$\psi(x)=y_0(x)+\delta y_1(x)+{\rm O}\big(\delta^2\big).$$

Because we are treating $\delta$ as small, the Stokes wedges have an angular
opening close to $2\pi/3$ and are approximately centered about the angles
$\theta_{\rm L}=-7\pi/6$ and $\theta_{\rm R}=\pi/6$. We construct solutions
$\psi_{\rm L}(x)$ and $\psi_{\rm R}(x)$ in the left and right Stokes wedges. We
then patch together these eigenfunctions and their first derivatives at the
origin $x=0$. The patching condition is
\begin{equation}
0=\psi_{\rm R}(x)\psi_{\rm L}'(x)-\psi_{\rm L}(x)\psi_{\rm R}'(x)\bigg|_{x=0}.
\label{e6}
\end{equation}

To zeroth order in powers of $\delta$ the Schr\"odinger eigenvalue equation
$H\psi(x)=E\psi(x)$ reads
$$y_0''(x)+ixy_0(x)+Ey_0(x)=0.$$
Substituting $x=re^{i\theta_{\rm L,R}}$ reduces this equation to an Airy
equation \cite{R24} for the zeroth-order eigenfunctions $y_{0,{\rm(L,R)}}(x)$
in the left and right wedges:
\begin{equation}
y''_{0,{\rm(L,R)}}(r)-\left(r-Ee^{\mp i\pi/3}\right)y_{0,{\rm(L,R)}}(r)=0,
\label{e7}
\end{equation}
where the derivatives are now taken with respect to $r$.

The boundary conditions on the eigenfunctions in each wedge require that
$y_{0,{\rm(L,R)}}(r)\to0$ as $r\to\infty$, so the solutions to (\ref{e7})
are Airy functions \cite{R24}:
\begin{eqnarray}
y_{0,{\rm(L,R)}}(x)&=&C_{\rm L,R}(x){\rm Ai}\left(r-Ee^{\mp i\pi/3}\right)
\nonumber\\
&=&C_{\rm L,R}{\rm Ai}\left(\mp xe^{\pm i\pi/6}+Ee^{\pm 2i\pi/3}\right),
\label{e8}
\end{eqnarray}
where $C_{\rm L,R}$ are multiplicative constants.

The right side of the patching condition (\ref{e6}) for the zeroth-order
solutions is calculated from the Wronskian identity for Airy functions
\cite{R24}:
\begin{eqnarray}
&&\psi_{0,{\rm R}}(x)\psi_{0,{\rm L}}'(x)-\psi_{0,{\rm L}}(x)\psi_{0,{\rm
R}}'(x)\big|_{x=0}\nonumber\\
&&\quad=-C_{\rm L}C_{\rm R}\left[e^{-i\pi/6}{\rm Ai}\left(Ee^{-2i\pi/3}\right)
Ai'\left(Ee^{2i\pi/3}\right)\right.\nonumber\\
&&\qquad+\left.e^{i\pi/6}{\rm Ai}\left(Ee^{2i\pi/3}\right)
Ai'\left(Ee^{-2i\pi/3}\right)\right]\nonumber\\
&&\quad=-iC_{\rm L}C_{\rm R}{\rm W}\left[{\rm Ai}\left(Ee^{2i\pi/3}\right),
{\rm Ai}\left(Ee^{-2i\pi/3}\right)\right]\nonumber\\
&&\quad=\textstyle{\frac{1}{2\pi}}C_{\rm L}C_{\rm R}\neq0.
\label{e9}
\end{eqnarray}
When $\delta$ is exactly $0$, the potential is linear in $x$ and $y_{0,{\rm(L,R)
}}(x)$ are the {\it exact} solutions to the Schr\"odinger equation. The above
calculation shows that these solutions cannot be patched, and thus there are no
eigenvalues at all when $\vep=-1$ ($\delta=0$). This conclusion is consistent
with Fig.~\ref{F3}, which shows that the real parts of all of the eigenvalues
become infinite as $\vep$ approaches $-1$. The fact that the spectrum is empty
at $\vep=-1$ is not a new result; the absence of eigenvalues of a linear
potential was established in Ref.~\cite{R25}.

Next, we perform a first-order ${\rm O}\big(\delta^1\big)$ analysis. We set
$y_1(x)=Q(x)y_0(x)$. (This substitution is motivated and explained in detail in
Ref.~\cite{R21}.) The first-order Schr\"odinger equation now reads
$$y_1''(x)+ixy_1(x)+ix\ln(ix)y_0(x)+Ey_1(x)=0.$$
We multiply this equation by the integrating factor $y_0(x)$ and insert the
leading-order approximation to the eigenfunctions and obtain
$$\left[ y_0^2(x) Q'(x)\right]' = -ix\ln(ix) y_0^2(x).$$
We then integrate this equation along the center ray of each Stokes wedge:
\begin{eqnarray}
Q'_{\rm L,R}(x)&=& i\int_0^{\mp\exp(\mp i\pi/6)\infty} dt\,t\ln(it)\left[\frac{
y_{0,{\rm(L,R)}}(t)} {y_{0,{\rm(L,R)}}(x)}\right]^2\nonumber\\
&=& ie^{\mp i\pi/3}\int_0^\infty ds\,s\ln\left(\mp s e^{\mp i\pi/6}\right)
\nonumber\\
&&\quad\times\left[\frac{y_{0,{\rm(L,R)}}\big(\mp se^{\mp i\pi/6}\big)}
{y_{0,{\rm(L,R)}}(x)}\right]^2\nonumber\\
&=&ie^{\mp i\pi/3}\int_0^\infty ds\,s\ln\left(se^{\pm 2i\pi/3}\right)\nonumber\\
&&\quad\times\left[\frac{{\rm Ai}\left(s+Ee^{\pm2i\pi/3}\right)}{{\rm Ai}\left(
\mp xe^{\pm i\pi/6}+Ee^{\pm 2i\pi/3}\right)}\right]^2.
\label{e10}
\end{eqnarray}

Thus, to first order in $\delta$ with $\psi(x)=y_0(x)[1+\delta Q(x)]$
the patching condition (\ref{e6}) becomes
\begin{eqnarray}
0&=&\left[1+\delta Q_{\rm R}(0)+\delta Q_{\rm L}(0)\right]\left[ y_{\rm 0,R}(x)
y_{\rm 0,L}'(x)\right.\nonumber\\
&&\quad\left.-y_{\rm 0,L}(x)y_{\rm 0,R}'(x)\right]_{x=0}\nonumber\\
&&\quad+\delta y_{\rm 0,L}(x)y_{\rm 0,R}(x)\left[Q_{\rm L}'(0)-Q_{\rm R}'(0)
\right] \nonumber\\
&=& C_{\rm L}C_{\rm R}\left\{-\textstyle{\frac{1}{2\pi}}+\delta{\rm Ai}\big(E
e^{-2\pi i/3}\big){\rm Ai}\big(Ee^{2\pi i/3}\big)\right.\nonumber\\
&&\left.\quad\times\left[Q_{\rm L}'(0)-Q_{\rm R}'(0)\right]\right\},
\label{e11}
\end{eqnarray}
where we have used the zeroth-order patching condition (\ref{e6}) and the
leading-order eigenfunction (\ref{e8}). Note that because the Schr\"odinger
equation is linear we are free to choose $Q_{\rm L}(0)+Q_{\rm R}(0)=0$.

For large $E$, we use the asymptotic expansion of the Airy function \cite{R24}
$${\rm Ai}(x)\textstyle{\sim\frac{1}{2\sqrt{\pi}}}x^{-1/4}\exp\left(-
\textstyle{\frac{2}{3}}x^{3/2}\right)\quad(|x|\to\infty,~|{\rm arg}\, x|<\pi).$$
Thus, the patching condition for $|E|\to\infty$ becomes
\begin{equation}
\textstyle{\frac{2}{\delta}}\sim\textstyle{\frac{1}{\sqrt{E}}}\exp\left(
\fourthird E^{3/2}\right)\left[Q_{\rm R}'(0)-Q_{\rm L}'(0)\right].
\label{e12}
\end{equation}
Note that because we are treating $\delta$ as small, the difference $Q_{\rm R}'
(0)-Q_{\rm L}'(0)$ is approximately a positive real number. For real $E$ this
difference is exactly real because $Q_{\rm R}'(0)$ and $-Q_{\rm L}'(0)$ are
complex conjugates.

We expand the right side of (\ref{e12}) to first order in $\beta/\alpha$, where 
$\alpha={\rm Re}\,E>0$ and $\beta={\rm Im}\,E$. This expansion is justified
because, as we can see in Fig.~\ref{F3}, the imaginary parts are small compared
with the real parts near $\vep=-1$. The patching condition (\ref{e12}) then
becomes
\begin{eqnarray}
\textstyle{\frac{2}{\delta}}&\sim&\alpha^{-1/2}\left(1+i\textstyle{\frac{\beta}{
\alpha}}\right)^{-1/2}\exp\left[\fourthird\alpha^{3/2}\left(1+i\textstyle{\frac{
\beta}{\alpha}}\right)^{3/2}\right]\nonumber\\
&=&\alpha^{-1/2}\left(1-i\textstyle{\frac{\beta}{2\alpha}}\right)\exp\left(
\fourthird\alpha^{3/2}\right)\exp\left(-2i\alpha^{1/2}\beta\right)\nonumber\\
&&\quad+{\rm O}\left(\textstyle{\frac{\beta^2}{\alpha^2}}\right).
\label{e13}
\end{eqnarray}

Hence, when $\delta$ is positive, we obtain the condition
$${\rm arg}\textstyle{\frac{2}{\delta}}={\rm arctan}\left(-\textstyle{\frac{
\beta}{2\alpha}}\right)-2\alpha^{1/2}\beta=2m\pi,$$
where $m$ is an integer. This result simplifies because the arctangent term
is small; to leading-order we obtain $2\alpha^{1/2}\beta=2m\pi$.
Similarly when $\delta<0$, we find that $2\alpha^{1/2}\beta=(2m+1)\pi$.

We conclude that for either sign of $\delta$ we obtain a simple formula for the
real part of the eigenvalues. Specifically, if we combine the above three
equations, we obtain $\textstyle{\frac{2}{|\delta|}}\sim\alpha^{-1/2}\exp
\left(\fourthird\alpha^{3/2}\right)$.
Hence, in the neighborhood of $\vep=-1$ (that is, when $\delta$ is near
$0$), the real parts of the eigenvalues are logarithmically divergent while the
imaginary parts of the eigenvalues remain finite:
\begin{equation}
{\rm Re}\,E\sim\left(-\textstyle{\frac{3}{4}}\ln|\delta|\right)^{2/3},\qquad
{\rm Im}\,E\sim\textstyle{\frac{n\pi}{2\sqrt{{\rm Re}\,E}}},
\label{e14}
\end{equation}
where $n$ is an even integer for $\delta>0$ and $n$ is an odd integer for
$\delta<0$. Evidently, the imaginary parts of the eigenvalues vary rapidly as
$\vep$ passes through $-1$ because there is a {\it logarithmic} singularity at
$\vep=-1$. A blow-up of the region $-1.05<\vep<1.05$ is given in Fig.~\ref{F4}.

\begin{widetext}

\begin{figure}[b!]
\begin{center}
\includegraphics[scale=0.21]{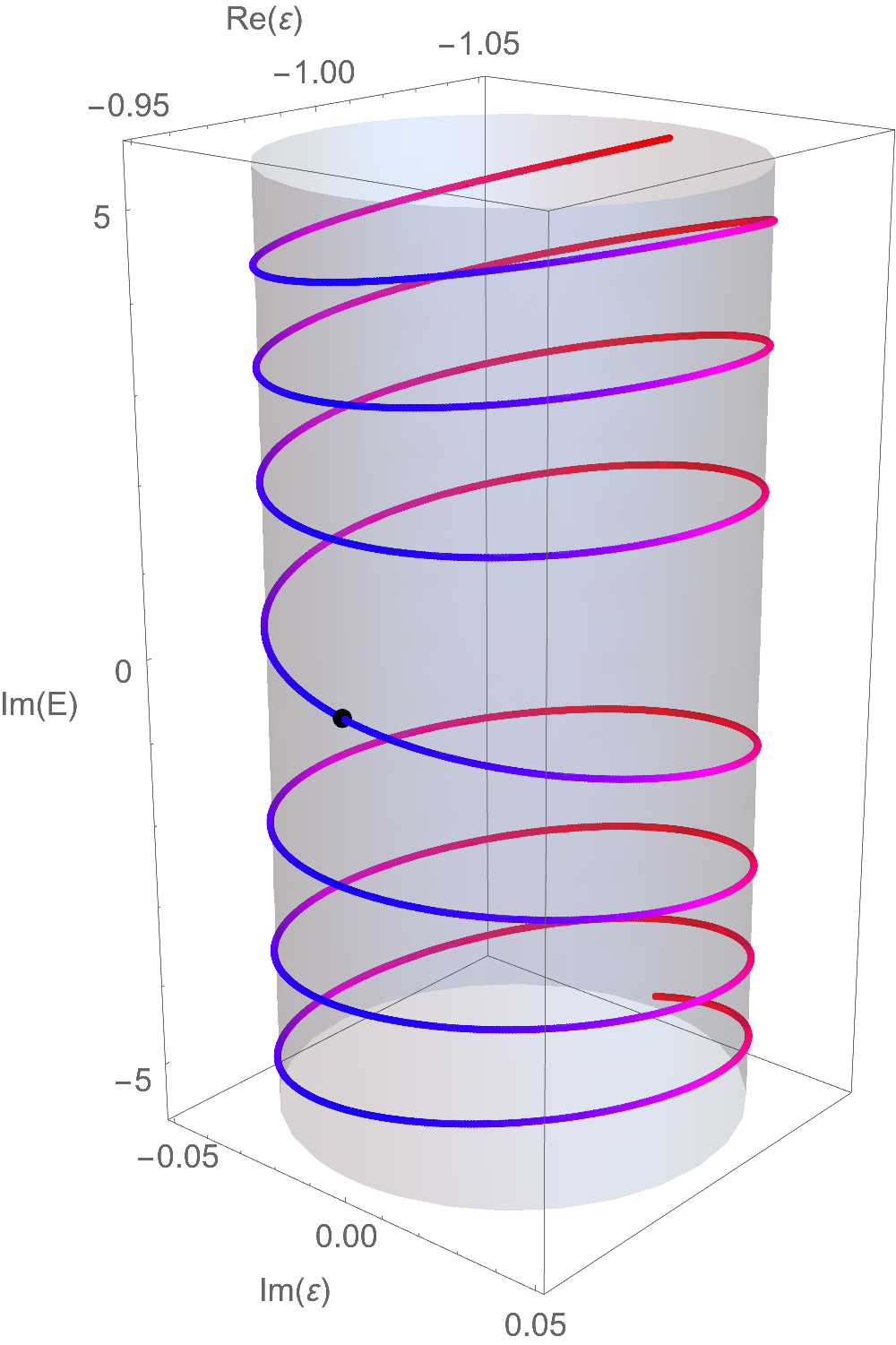}\hspace{.5cm}
\includegraphics[scale=0.225]{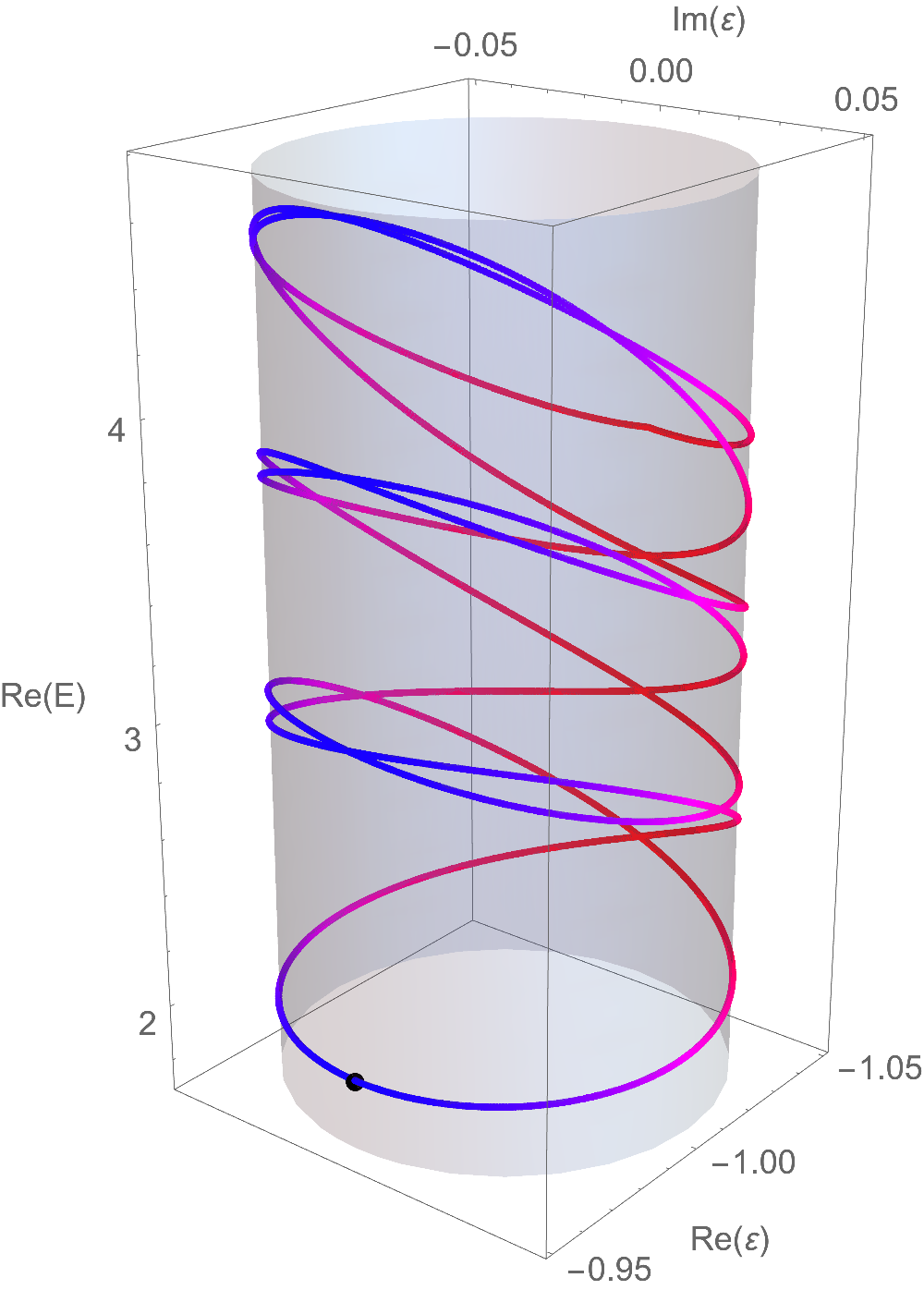}
% \includegraphics[trim=0mm 0mm 0mm 0mm,clip=true,scale=0.35]{Fig5.png}
% trim=left bottom right top
\end{center}
\caption{[Color online] Behavior of the eigenvalues of the Hamiltonian $H=p^2+
x^2(ix)^\vep$ as the parameter $\vep$ winds around the exceptional point at
$\vep=-1$ in a circle of radius $0.05$ in the complex-$\vep$ plane. This
singular point is an infinite-order exceptional point, and all of the complex
eigenvalues analytically continue into one another as one encircles the
exceptional point. The lines are shaded blue when ${\rm Re}\,\vep>0$ and red
when ${\rm Im}\,\vep<0$. The behavior of the imaginary parts of the
eigenvalues (left panel) are easier to visualize because they exhibit a simple
logarithmic spiral. The dot shows that the imaginary part of an eigenvalue (the
eigenvalue shown in black in Figs.~\ref{F3} and \ref{F4}) vanishes (the
eigenvalue is real) when ${\rm Re}\,\vep>0$. However, as we wind in one
direction the imaginary parts of the eigenvalues increase in a helical fashion
and as we wind in the opposite direction the imaginary parts of the eigenvalues
decrease in a helical fashion. As we pass the real-$\vep$ axis we pass through
the values plotted on the red dashed lines shown in Figs.~\ref{F3} and
\ref{F4}. A shaded cylinder has been drawn to assist the eye in the following
this helix. The behavior of the real parts of the eigenvalues (right panel) is
more complicated because the curves form a {\it double} helix. The two helices
intersect {\it four} times each time the singular point at $\vep=-1$ is
encircled, and they intersect at $90^\circ$ intervals. If we begin at the dot,
we see that the real parts of the eigenvalues increase as we rotate about
$\vep=-1$ in either direction. Each time $\vep$ crosses the real axis in the
complex-$\vep$ plane the curves pass through the values shown at the left and
right edges of Fig.~\ref{F4}.}
\label{F5}
\end{figure}

% \vspace{1.5cm}
\end{widetext}

To visualize the behavior of the eigenvalues near $\vep=-1$ more clearly, we
have plotted the imaginary and real parts of the eigenvalues in the {\it complex
}-$\vep$ plane in the left and right panels of Fig.~\ref{F5}. Observe that the
imaginary parts of the eigenvalues lie on a helix and that the real parts of the
eigenvalues lie on a {it double} helix as $\vep$ winds around the logarithmic
singularity at $\vep=-1$. This logarithmic singularity is an {\it
infinite-order} exceptional point, which one discovers only very rarely in
studies of the analytic structure of eigenvalue problems.

\section{Eigenvalue behavior as $\vep\to-2$}
\label{s3}
\begin{figure}[t!]
\begin{center}
\includegraphics[trim=0mm 0mm 0mm 0mm,clip=true,scale=0.41]{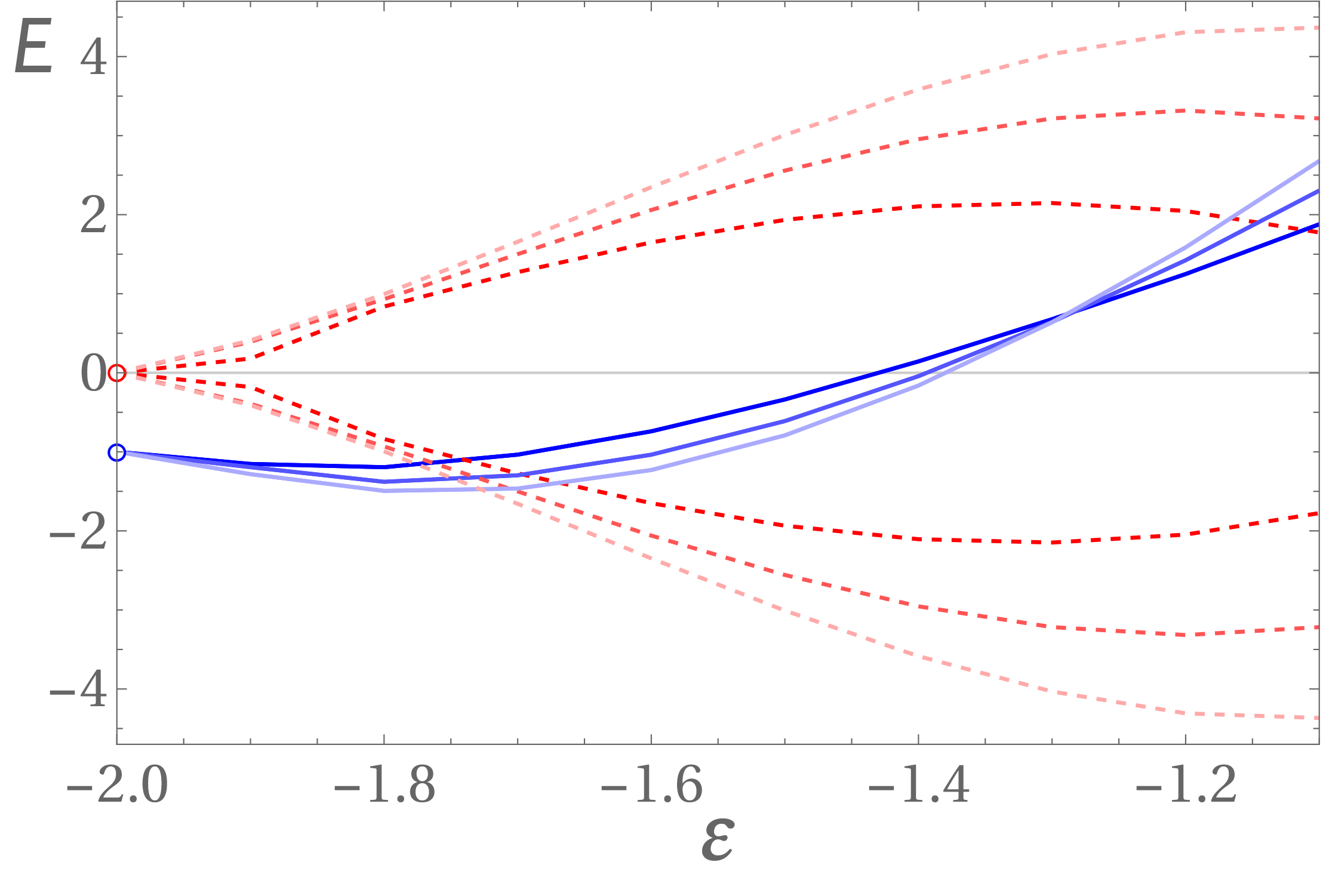}
% trim=left bottom right top
\end{center}
\caption{[Color online] First three complex-conjugate pairs of eigenvalues of
the Hamiltonian $H=p^2+x^2(ix)^\vep$ plotted as functions of $\vep$ for $-2\leq
\vep\leq-1.1$. This figure is a continuation of Fig.~\ref{F3}. Note that the
real parts of the eigenvalues coalesce to $-1$ and the imaginary parts coalesce
to 0 as $\vep$ approaches $-2$. The results of a WKB calculation of these
eigenvalues near $\vep=-2$ is given in (\ref{e29}). Note that the real parts of
the eigenvalues cross near $\vep=-1.3$, but they do not all cross at the same
point as can be seen in Fig.~\ref{F7}.}
\label{F6}
\end{figure}

\begin{figure}[t!]
\begin{center}
\includegraphics[scale=0.45]{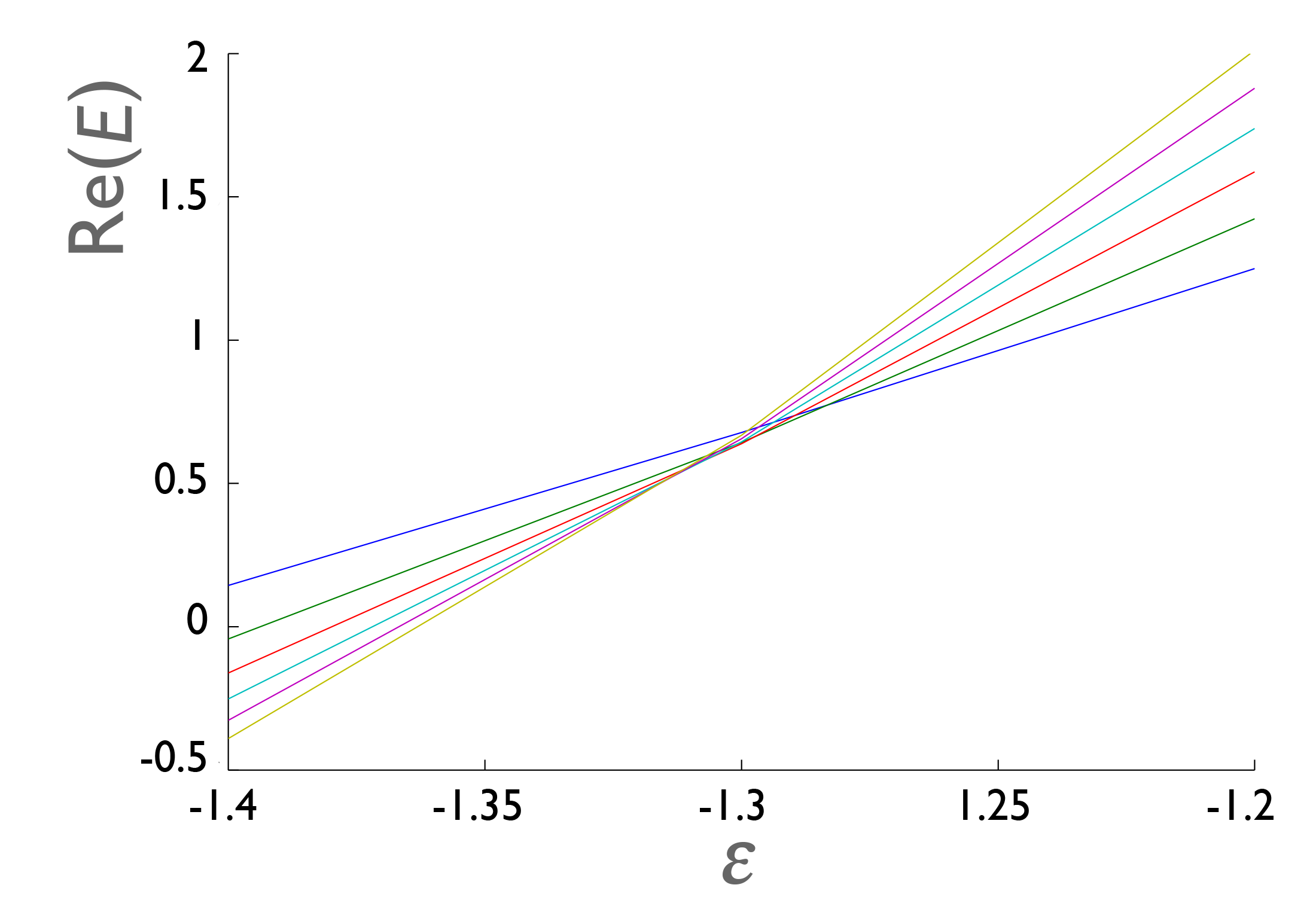}
\end{center}
\caption{[Color online] Detail of Fig.~\ref{F6} showing the behavior of the real
parts of the first six eigenvalues of the Hamiltonian $H=p^2+x^2(ix)^\vep$ for
$-1.4\leq\vep\leq-1.2$. The real parts of the eigenvalues cross almost at the
same value of $\vep$ but the imaginary parts of the eigenvalues remain well
separated.}
\label{F7}
\end{figure}

In Fig.~\ref{F6} we plot the first three complex-conjugate pairs of eigenvalues
in the range $-2.0\leq\vep\leq-1.1$. Note that the eigenvalues $E_k$ coalesce to
the value $-1$ as $\vep$ approaches $-2$. As $\vep$ decreases towards $-2$ the
real part of $E_k$ becomes more negative as $k$ increases, and the spectrum
becomes {\it inverted}; that is, the higher-lying real parts of the eigenvalues
when $\vep$ is near $-1.7$ (for example) decrease as $\vep$ decreases and they
cross when $\vep$ is near $-1.3$. This crossing region is shown in
detail in Fig.~\ref{F7}.

\clearpage

The objective of this section is to explain the behavior of the eigenvalues as
$\vep$ approaches $-2$ by performing a local analysis near $\vep=-2$. To do so
we let
$$\vep=-2+\delta$$
and treat $\delta$ as small ($\delta<<1$) and positive. With this change of
parameter (\ref{e2}) becomes
\begin{equation}
-y''(x)-(ix)^\delta y(x)=Ey(x).
\label{e15}
\end{equation}
The boundary conditions on $y(x)$, which we can deduce from Fig.~\ref{F2}, are
that the eigenfunctions $y(x)$ must vanish asymptotically at the ends of a path
that originates at $e^{-3\pi i/2}\infty$ in the complex-$x$ plane, goes down to
the origin along the imaginary axis, encircles the origin in the positive
direction, goes back up the imaginary axis, and terminates at $e^{\pi i/2}
\infty$. The eigenfunctions are required to vanish at the endpoints $e^{-3\pi i/
2}\infty$ and $e^{\pi i/2}\infty$.

We now make the crucial assumption that it is valid to expand the potential term
in (\ref{e15}) as a series in powers of $\delta$. To second order in $\delta$ we
then have
\begin{eqnarray}
&&-y''(x)-\delta \ln(ix)y(x)-\half\delta^2[\ln(ix)]^2y(x)\nonumber\\
&&\qquad\qquad=(E+1)y(x).
\label{e16}
\end{eqnarray}
In this form one can see that to every order in powers of $\delta$ the potential
terms in the Schr\"odinger equation are singular at $x=0$. As a consequence, the
solution $y(x)$ vanishes at $x=0$. (One can verify that $y(0)=0$ by examining
the WKB approximation to $y(x)$; the prefactor $[V(x)-E]^{-1/4}$ vanishes
logarithmically.)

We then make the change of independent variable $t=-ix$. In terms of $t$
(\ref{e16}) becomes
\begin{eqnarray}
&&-y''(t)+\delta\ln(-t)y(t)+\half\delta^2[\ln(-t)]^2y(t)\nonumber\\
&&\qquad\qquad=-(E+1)y(t).
\label{e17}
\end{eqnarray}
This eigenvalue equation is posed on a contour on the real-$t$ axis that
originates at $t=+\infty$, goes down the positive-real $t$ axis, encircles the
origin in the positive direction, and goes back up to $e^{2\pi i}\infty$, and
$y(t)$ is required to vanish at the endpoints of this contour. We then replace
$\ln(-t)$ with $\ln(t)\pm i\pi$:
\begin{eqnarray}
&&-y''(t)+\delta[\ln(t)\pm i\pi]y(t)+\half\delta^2[\ln(t)\pm i\pi]^2y(t)
\nonumber\\
&&\qquad\qquad=-(E+1)y(t).
\label{e18}
\end{eqnarray}

Next, we make the scale change
$$t=s/\sqrt{\delta}.$$
This converts (\ref{e18}) into the Schr\"odinger equation
\begin{equation}
-y''(s)+\ln(s)y(s)+\delta U(s)y(s)=Fy(s),
\label{e19}
\end{equation}
where the energy term $F$ is given by
\begin{eqnarray}
F&=&-(E+1)/\delta+\half\ln(\delta)\mp i\pi-\eighth\delta[\ln(\delta)]^2
\nonumber\\
&&\qquad+\half\delta\pi^2\pm\half\delta i\pi\ln(\delta)
\label{e20}
\end{eqnarray}
and the order $\delta$ term in the potential is given by
\begin{equation}
U(s)=\half[\ln(s)]^2-\half\ln(\delta)\ln(s)\pm i\pi\ln(s).
\label{e21}
\end{equation}

Our procedure will be as follows. First, we neglect the $U(s)$ term in
(\ref{e19}) because $\delta$ is small and we use WKB theory to solve the simpler
Schr\"odinger equation 
\begin{equation}
-y_0''(s)+\ln(s)y_0(s)=F_0 y_0(s).
\label{e22}
\end{equation}
Second, we find the energy shift $\Delta F$ due to the $U(s)$ term in
(\ref{e19}) by using first-order Rayleigh-Schr\"odinger theory \cite{R21}; to
wit, we calculate the expectation value of $U(s)$ in the WKB approximation to
$y_0(s)$ in (\ref{e22}). Having found $F=F_0+\Delta F$, we obtain the energy $E$
from (\ref{e20}):
\begin{eqnarray}
E&=&-1-F\delta+\half\delta\ln(\delta)\mp i\pi\delta-\eighth\delta^2
[\ln(\delta)]^2\nonumber\\
&&\quad+\half\pi^2\delta^2\pm\half i\pi\delta^2\ln(\delta).
\label{e23}
\end{eqnarray}
This approach gives a very good numerical approximation to the energies shown in
Fig.~\ref{F6}.

The standard WKB quantization formula for the eigenvalues $F_0$ in a single-well
potential $V(s)$ (the two-turning-point problem) is 
\begin{equation}
\left(n+\half\right)\pi=\int_{s_1}^{s_2}ds\sqrt{F_0-V(s)}\quad(n>>1).
\label{e24}
\end{equation}
For (\ref{e22}) the potential $V(s)$ is $\ln(s)$ and the boundary conditions on
$y_0(s)$ are given on the positive half line: $y_0(s)$ vanishes at $s=0$ and at
$s=+\infty$. In order to apply (\ref{e24}) we extend the differential equation
to the whole line $-\infty<s<+\infty$ by replacing $\ln(s)$ with $\ln(|s|)$
and consider only the {\it odd-parity} solutions. Thus, we must replace the
integer $n$ in (\ref{e24}) with $2k+1$, where $k=0,\,1,\,2,\,\ldots$. The
turning points are given by $s_1=-e^{F_0}$ and $s_2=e^{F_0}$. Hence, the WKB
formula (\ref{e24}) becomes
\begin{eqnarray}
\left(2k+1+\half\right)\pi&=&\int_{-e^{F_0}}^{e^{F_0}}ds\sqrt{F_0-\ln(|s|)}
\nonumber\\
&=&2\int_0^{e^{F_0}}ds\sqrt{F_0-\ln(s)}\quad(k>>1).\nonumber
\end{eqnarray}
The substitution $s=ue^{F_0}$ simplifies this equation to
$$\left(2k+\threehalf\right)\pi=2e^{F_0}\int_0^1 du\sqrt{-\ln(u)}$$
and the further substitution $v=-\ln(u)$ reduces the integral to a Gamma
function:
$$\int_0^1 du\sqrt{-\ln(u)}=\int_0^\infty dv\,e^{-v}v^{1/2}=\Gamma\left(
\threehalf\right)=\half\sqrt{\pi}.$$
Thus, the WKB approximation to the eigenvalues $F_0$ is
\begin{equation}
F_0=\ln\left[\left(2k+\threehalf\right)\sqrt{\pi}\right],
\label{e25}
\end{equation}
which is valid for large $k$.

Next, we calculate the order-$\delta$ correction $\Delta F$ to (\ref{e25}) due
to the potential $U(s)$ in (\ref{e19}). To do so we calculate the expectation
value of $U(s)$ in the WKB eigenfunction $y_0(s)$ of (\ref{e22}):
\begin{equation}
\Delta F=\delta\int_0^\infty ds\,U(s)[y_0(s)]^2\bigg/\int_0^\infty ds\,[y_0(s)
]^2,
\label{e26}
\end{equation}
where $U(s)$ is given in (\ref{e21}).

Integrals of this type are discussed in detail in Chap.~9 of Ref.~\cite{R21}. To
summarize the procedure, in the classically-forbidden region beyond the turning
point, $y_0(s)$ is exponentially small, and the contribution to the integral
from this region is insignificant. In the classically-allowed region the square
of the eigenfunction has the general WKB form 
$$[y_0(s)]^2=\frac{C}{\sqrt{F_0-V(s)}}\sin^2\left[\phi+\int^s dr\,
\sqrt{F_0-V(r)}\right],$$
where $C$ is a multiplicative constant and $\phi$ is a constant phase shift.

Making the replacement $\sin^2\theta=\half-\half\cos(2\theta)$, we observe that
because of the Riemann-Lebesgue lemma, the cosine term oscillates to zero for
large quantum number $k$, and we may replace $[y_0(s)]^2$ in the integrals in
(\ref{e26}) by the simple function $\half[F_0-V(s)]^{-1/2}$. Thus, the shift in
the eigenvalues is given by
\begin{eqnarray}
\Delta F&=&\delta\int_0^{e^{F_0}}{\frac{ds\,\ln(s)}{\sqrt{F_0-\ln(s)}}}\left[
\half\ln(s)-\half\ln(\delta)\pm i\pi\right]\nonumber\\
&&\bigg/\int_0^{e^{F_0}}\frac{ds}{\sqrt{F_0-\ln(s)}}.
\label{e27}
\end{eqnarray}
After making the previous changes of variable $s=e^{F_0}u$ followed by $v=-
\ln(u)$, we obtain
\begin{eqnarray}
\Delta F&=&\delta\int_0^\infty dv\,e^{-v}(F_0-v)v^{-1/2}\left[\half(F_0-v)
\right.\nonumber\\
&&-\left.\half\ln(\delta)\pm i\pi\right]\bigg/\int_0^\infty dv\,e^{-v}v^{-1/2},
\nonumber
\end{eqnarray}
which evaluates to
\begin{eqnarray}
\Delta F&=&\eighth\delta\left[4F_0^2-4F_0+3-4F_0\ln(\delta)\right.\nonumber\\
&&\qquad\left.+2\ln(\delta)\pm i\pi(8F_0-4)\right].
\label{e28}
\end{eqnarray}
Finally, we substitute $F=F_0+\Delta F$ in (\ref{e23}) to obtain the
eigenvalues $E_k$:
\begin{eqnarray}
E_k&=&-1+\delta\left[\half\ln(\delta)-F_0\right]-\eighth\delta^2\left\{
[\ln(\delta)]^2+2\ln(\delta)\right.\nonumber\\
&& \left.-4\ln(\delta)F_0+3-4\pi^2-4F_0+4F_0^2\right\} \nonumber\\
&&\pm i\left\{-\delta\pi+\half\delta^2\left[\pi\ln(\delta)+\pi-2F_0\right]
\right\},
\label{e29}
\end{eqnarray}
where $F_0$ is given in (\ref{e25}).

To verify these results, in Tables \ref{t1} and \ref{t2} we compare our
numerical calculation of ${\rm Re}\,E_k$ and ${\rm Im}\,E_k$ with the
asymptotic prediction in (\ref{e29}).

\begin{table}[h!]
\begin{center}
\begin{tabular}{|c|c|c|c|}
\hline
$k$ & Numerical value of & ${\rm O}(\delta^2)$ calculation & relative\\
& ${\rm Re}\,E_k$ at $\delta=0.01$ & of ${\rm Re}\,E_k$ in (\ref{e29}) & error\\
\hline
0  & $-1.0352$ & $-1.0414$ & $8.70\,\%$ \\
2  & $-1.0426$ & $-1.0461$ & $0.33\,\%$ \\
4  & $-1.0469$ & $-1.0493$ & $0.30\,\%$ \\
6  & $-1.0499$ & $-1.0518$ & $0.18\,\%$ \\
8  & $-1.0523$ & $-1.0538$ & $0.15\,\%$ \\
10 & $-1.0542$ & $-1.0555$ & $0.12\,\%$ \\
12 & $-1.0559$ & $-1.0569$ & $0.10\,\%$ \\
\hline
\end{tabular}
\end{center}
\caption{\label{t1} Comparison of the real parts of the eigenvalues of the
differential equation (\ref{e15}) at $\delta=0.01$ with the asymptotic
approximation in (\ref{e29}). The rate at which the accuracy increases with
increasing $k$ is similar to the increase in accuracy of the standard WKB
approximation to the eigenvalues of the quartic anharmonic oscillator 
\cite{R21}.}
\end{table}

\begin{table}[h!]
\begin{center}
\begin{tabular}{|c|c|c|c|}\hline
$k$ & Numerical value of & ${\rm O}(\delta^2)$ calculation & relative\\
& ${\rm Im}\,E_k$ at $\delta=0.01$ & of ${\rm Im}\,E_k$ in (\ref{e29}) & error\\
\hline
0  & $0.03397$ & $0.03210$ & $5.3\,\%$ \\
2  & $0.03352$ & $0.03220$ & $3.8\,\%$ \\
4  & $0.03339$ & $0.03224$ & $3.4\,\%$ \\
6  & $0.03334$ & $0.03226$ & $3.2\,\%$ \\
8  & $0.03332$ & $0.03228$ & $3.1\,\%$ \\
10 & $0.03332$ & $0.03229$ & $3.0\,\%$ \\
12 & $0.03333$ & $0.03231$ & $3.0\,\%$ \\
\hline
\end{tabular}
\end{center}
\caption{\label{t2} Comparison of the imaginary parts of the eigenvalues of the
differential equation (\ref{e15}) at $\delta=0.01$ with the asymptotic
approximation in (\ref{e29}).}
\end{table}

\section{Eigenvalue behavior for $-4<\vep<-2$}
\label{s4}

This section reports our numerical calculations of the eigenvalues for $\vep$
between $-2$ and $-4$. We rotate $x$ in
(\ref{e2}) by $90^\circ$ by making the transformation $s=ix$. In the $s$
variable the eigenvalue equation (\ref{e2}) becomes 
\begin{equation}
\psi''(s)-s^{2+\vep}\psi(s)=E\psi(s).
\label{e30}
\end{equation}
In the $x$ variable the center-of-wedge angles (\ref{e3}) are $-\pi+\vep\pi/(8
+2\vep)$ and $-\vep\pi/(8+2\vep)$ but in the $s$ variable these
angles are simply $\mp2\pi/(4+\vep)$. Thus, the integration contour makes $2/(
4+\vep)$ loops around the logarithmic branch-point at the origin in
the complex-$s$ plane.

\begin{widetext}

\begin{figure}[b!]
\begin{center}
\includegraphics[scale=0.44]{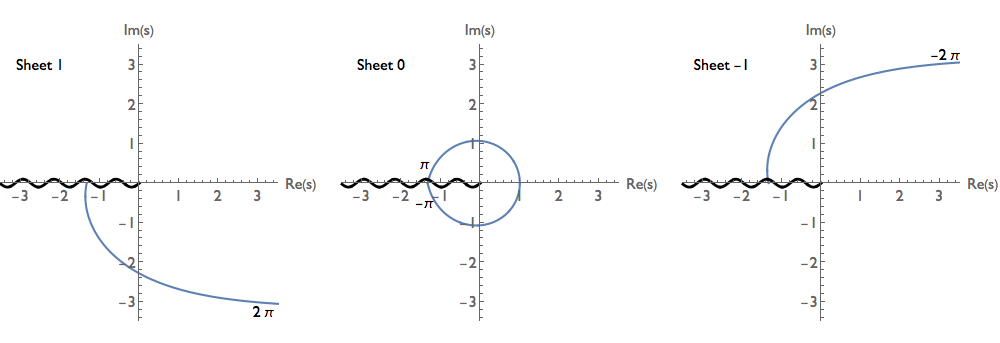}
\end{center}
\caption{[Color online] Contour in the complex-$s$ plane for the complex Coulomb
potential $\vep=-3$. The contour comes in from $\infty$ parallel to the
positive-real axis at an angle of $-2\pi$ in the center of the left Stokes wedge
(right panel). Next, it loops around the origin in the positive direction
(center panel). Finally, it goes back out to $\infty$ parallel to the
positive-real axis at an angle of $2\pi$ in the center of the right Stokes wedge
(left panel). The total rotation about the origin is $4\pi$.}
\label{F8}
\end{figure}

\end{widetext}

\begin{figure}[t!]
\begin{center}
\includegraphics[scale=0.24]{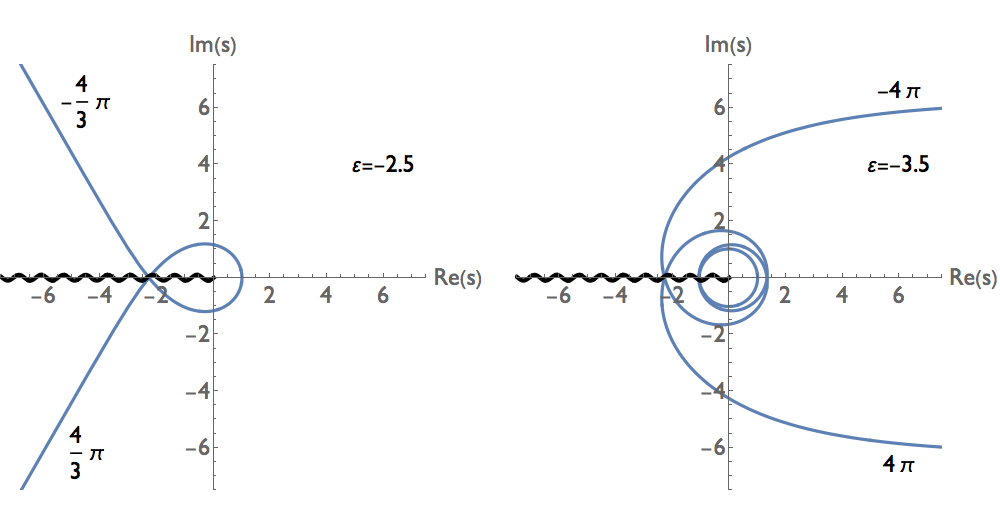}
\end{center}
\caption{[Color online] Eigenvalue contours in the complex-$s$ plane for the
cases $\vep=-2.5$ and $\vep=-3.5$.}
\label{F9}
\end{figure}

For example, if $\vep=-3$ (this is the complex $\cPT$-symmetric version of the
Coulomb potential for which $H=p^2+i/x$ \cite{R27}), the contour loops around
the origin exactly twice; it goes from an angle $-2\pi$ to the angle $2\pi$.
Looping contours for other complex eigenvalue problems have been studied in the
past and have been called ``toboggan contours'' \cite{R28}. In the
$\cPT$-symmetric Coulomb case the contour is shown in Fig.~\ref{F8}. Figure
\ref{F9} shows the contours for the cases $\vep=-2.5$ and $\vep=-3.5$.

To solve these eigenvalue problems with looping contours we introduce the change
of variable
\begin{equation}
s(t)=\frac{1}{1-t^2}\exp\frac{2\pi it}{4+\vep},
\label{e31}
\end{equation}
which parametrizes the looping path in the complex-$s$
plane in terms of the
real variable $t$. As $t$ ranges from $-1$ to $+1$, the path in the complex-$s$
plane comes in from infinity in the center of the left Stokes wedge, loops
around the logarithmic branch-point singularity at the origin,
and goes back out
to infinity in the center of the right Stokes wedge. In terms of the $t$
variable the eigenvalue equation (\ref{e30}) has the form
\begin{equation}
\frac{\psi''(t)}{[s'(t)]^2}-\frac{s''(t)}{[s'(t)]^3}\psi'(t)-[s(t)]^{2+\vep}
\psi(t)=E\psi(t),
\label{e32}
\end{equation}
where $\psi(t)$ satisfies $\psi(-1)=\psi(1)=0$.

To solve this eigenvalue problem we use the Arnoldi algorithm, which has
recently come available on Mathematica \cite{R29}. This algorithm finds
low-lying eigenvalues,
whether \hfill or \hfill not \hfill they \hfill are \hfill real. \hfill We \hfill apply \hfill the \hfill Arnoldi \hfill al-

\noindent
gorithm to (\ref{e32}) subject to the homogeneous Dirichlet boundary conditions
$\psi(-1+\eta)=\psi(1-\eta)=0$ and let $\eta\to0^+$. There are two possible
outcomes: (i) In this limit, some eigenvalues rapidly approach limiting values;
these eigenvalues belong to the discrete part of the spectrum. (ii) Other
eigenvalues become dense on curves in the complex plane as $\eta\to0^+$; these
eigenvalues belong to the continuous part of the spectrum.

\subsection{$\vep$ slightly below $-2$}
\begin{widetext}

\begin{figure}[b!]
\begin{center}
\includegraphics[scale=0.335]{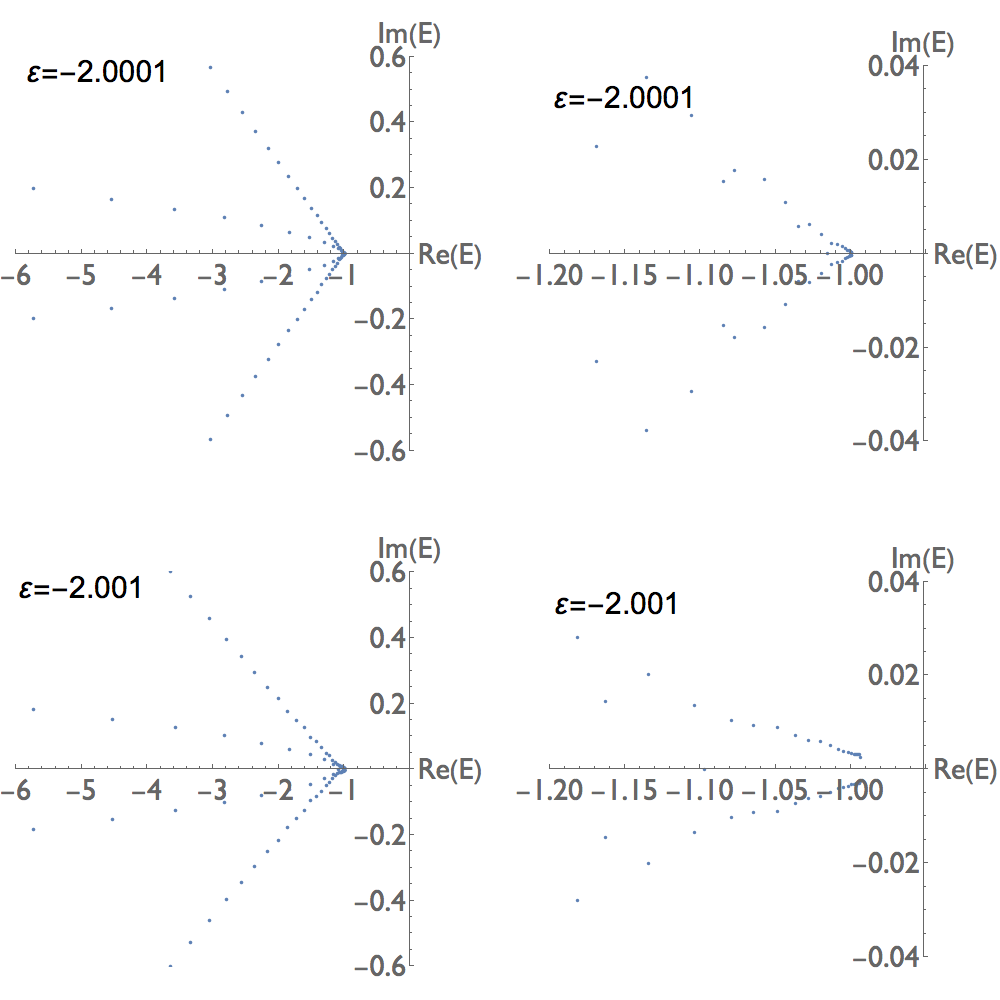}
\end{center}
\caption{[Color online] Eigenvalues of the Hamiltonian $H=p^2+x^2(ix)^\vep$ for
$\vep=-2.0001$ and $-2.001$. The spectrum lies in the left-half complex plane
and is partly continuous partly discrete. The eigenvalues in the continuous part
of the spectrum lie on a pair of complex-conjugate curves that radiate away from
$-1$ and as we calculate more eigenvalues, the points on these curves become
denser. The discrete part of the spectrum consists of eigenvalues lying on two
complex-conjugate curves that are much closer to the negative-real axis. There
is an elaborate structure near $\vep=-1$. Note that as $\vep$ goes below $-2$,
the eigenvalues move away from the point $-1$; specifically, for $\vep=-2.0001$
the distance from $-1$ to the nearest eigenvalue is about $0.0005$ and for $\vep
=-2.001$ the distance to the nearest eigenvalue is about $0.008$.}
\label{F10}
\end{figure}

\end{widetext}

As soon as $\vep$ goes below $-2$, the eigenvalues explode away from the value
$-1$ (shown at the left side of Fig.~\ref{F6}). In Fig.~\ref{F10} we plot about
100 eigenvalues for $\vep=-2.0001$
and \hfill $-2.001$. \hfill In \hfill each \hfill plot \hfill we \hfill see \hfill both \hfill discrete \hfill and \hfill con-

\noindent
tinuous eigenvalues. The continuous eigenvalues
lie on a complex-conjugate pair of curves in the left-half plane; the discrete
eigenvalues also lie in the left-half plane but closer to the real axis.

\subsection{Discrete and continuous eigenvalues}
\begin{figure}[t!]
\begin{center}
\includegraphics[scale=0.24]{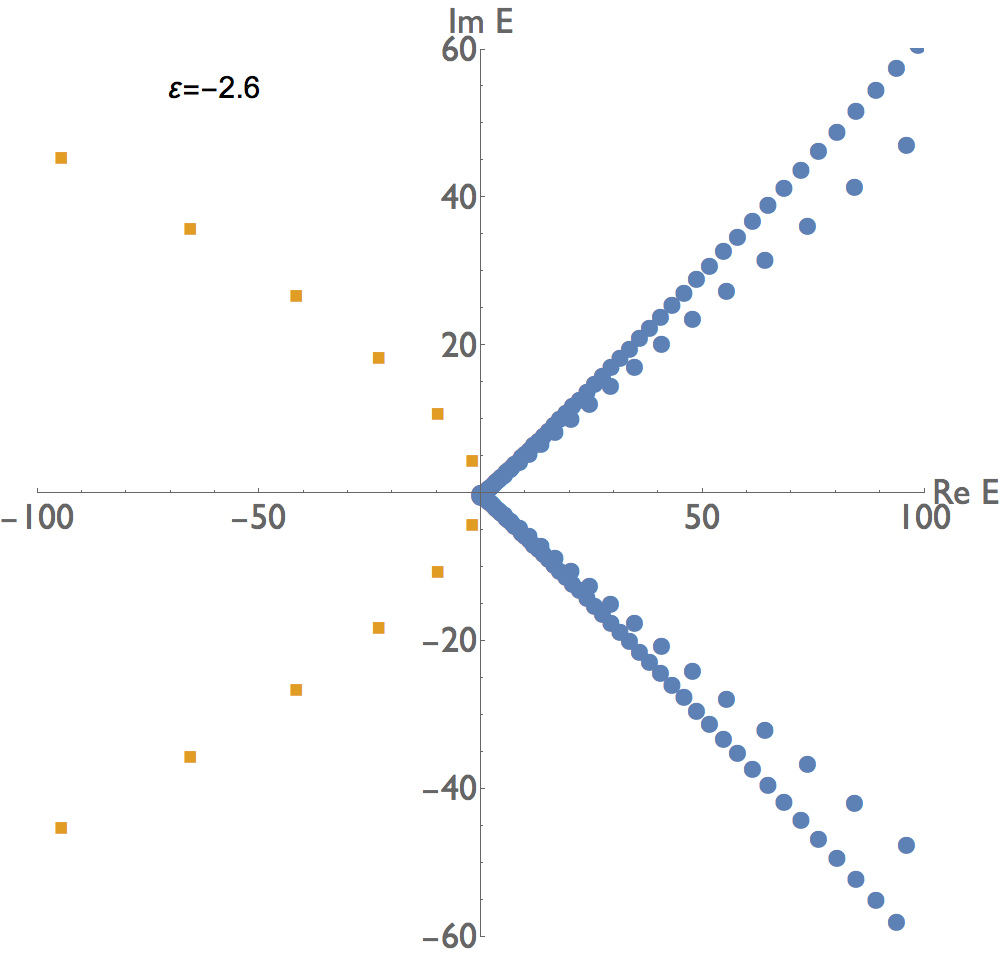}
\end{center}
\caption{[Color online] Discrete and continuous parts of the spectrum of the
$\cPT$-symmetric Hamiltonian $H=p^2+x^2(ix)^\vep$, for the case $\vep=-2.6$. The
discrete eigenvalues (orange squares) occur in pairs in the left-half complex
plane. The continuous eigenvalues (blue dots) lie on two complex-conjugate pairs
of curves in the right-half complex plane. As we decrease the cell size in
the Arnoldi algorithm, the dots become dense on these curves. The continuous
curves of eigenvalues originate slightly to the left of the origin.}
\label{F11}
\end{figure}

\begin{figure}[t!]
\begin{center}
\includegraphics[scale=0.22]{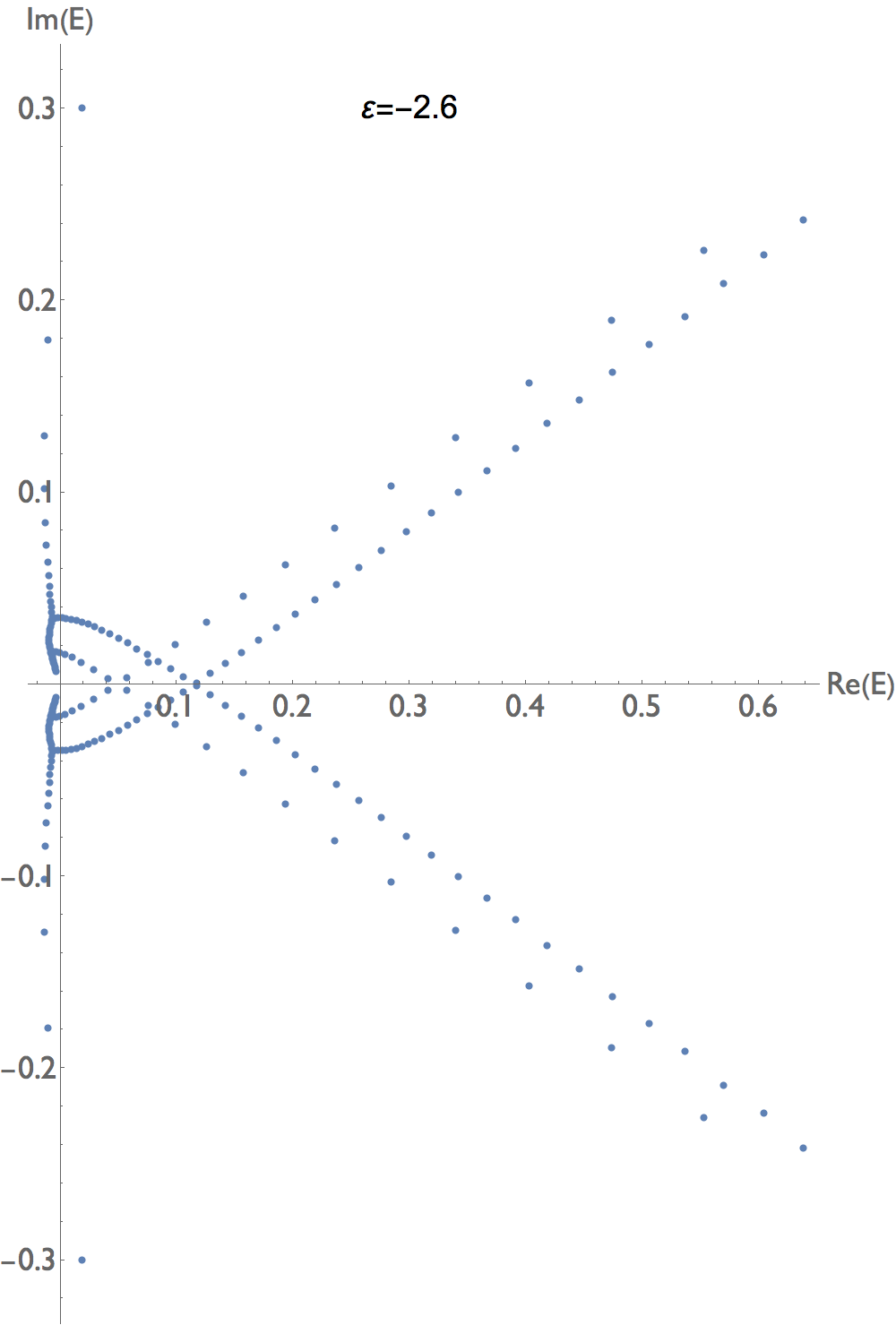}
\end{center}
\caption{[Color online] Detail of Fig.~\ref{F11} showing the elaborate
structure of the spectrum
near the origin in the complex-eigenvalue plane for $\vep=-2.6$.}
\label{F12}
\end{figure}

While the purpose of Fig.~\ref{F10} is to show that the eigenvalues explode
away from $-1$ as $\vep$ goes below $-2$, it is also important to show how
to distinguish between discrete and continuous eigenvalues. To illustrate
this we apply the Arnoldi algorithm at $\vep=-2.6$. Our results are given in
Fig.~\ref{F11} for $\eta=0.01$. The spectrum in this case is qualitatively
different from the spectrum near $\vep=-2$; there are now {\it two} pairs of
curves of continuous eigenvalues, and these curves are now in the right-half
complex
plane. The discrete eigenvalues are still in the left-half complex plane but
further from the negative real axis. There is an elaborate spectral structure
near the origin and this is shown in Fig.~\ref{F12}. (We do not investigate
this structure in this paper and reserve it for future research.) 

We emphasize that when the Arnoldi algorithm is used to study a spectrum, it
can only return discrete values. Thus, one must determine whether an
Arnoldi eigenvalue belongs to a discrete or a continuous part of the spectrum.
To distinguish between these two possibilities we study the associated
eigenfunctions and observe how they obey the boundary conditions. Plots of
discrete and continuous eigenfunctions associated with eigenvalues shown in
Fig.~\ref{F11} are given in Figs.~\ref{F13} and \ref{F14}.

In Fig.~\ref{F13} we plot the absolute values of the eigenfunctions
corresponding to the complex-conjugate pair of eigenvalues $E=-1.79\pm4.31i$
for $\vep=-2.6$. Observe that as $t$ approaches the boundaries $-1$ and $1$,
the eigenfunctions decay to $0$ exponentially. We conclude from this that
the eigenvalues are discrete. This result can then be verified by taking finer
cell sizes in the Arnoldi algorithm. As the cell size decreases,
the numerical values of $E$ are stable. In contrast, in Fig.~\ref{F14} in
which the absolute values of the eigenfunctions corresponding to the pair of
eigenvalues $E=-0.01\pm0.18i$ are plotted, we see that
the eigenfunctions vanish exponentially at one endpoint but vanish sharply
at the other endpoint. We therefore identify these eigenvalues as belonging
to the continuous spectrum. Decreasing the Arnoldi cell size results in a
denser set of eigenvalues along the same curve.

\subsection{Complex Coulomb potential $\vep=-3$}

\begin{figure}[t!]
\begin{center}
\includegraphics[scale=0.25]{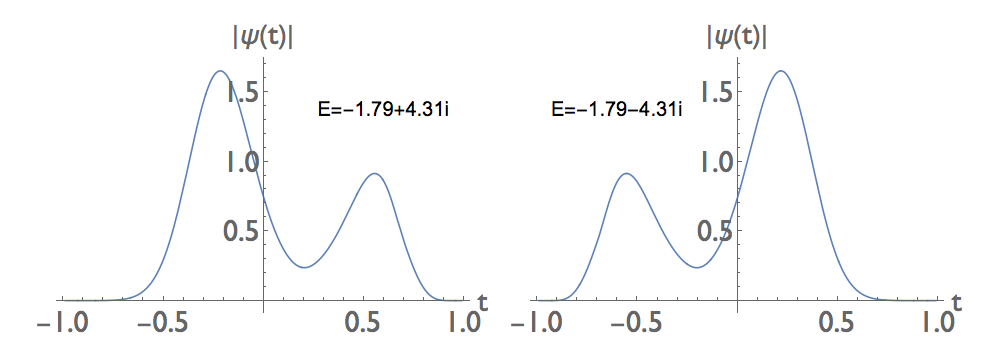}
\end{center}
\caption{[Color online] Absolute values of the eigenfunctions $\psi(t)$ for the
discrete eigenvalues $-1.79\pm4.31i$ for $\vep=-2.6$. The eigenfunctions satisfy
homogeneous boundary conditions at $\pm(1-\eta)$ for $\eta=0.01$
and look like bound-state eigenfunctions in the sense that the eigenfunctions
decay to $0$ exponentially fast at both boundary points. The left and right
panels are interchanged under $t\to-t$, which corresponds to a $\cPT$
reflection.}
\label{F13}
\end{figure}

\begin{figure}[t!]
\begin{center}
\includegraphics[scale=0.25]{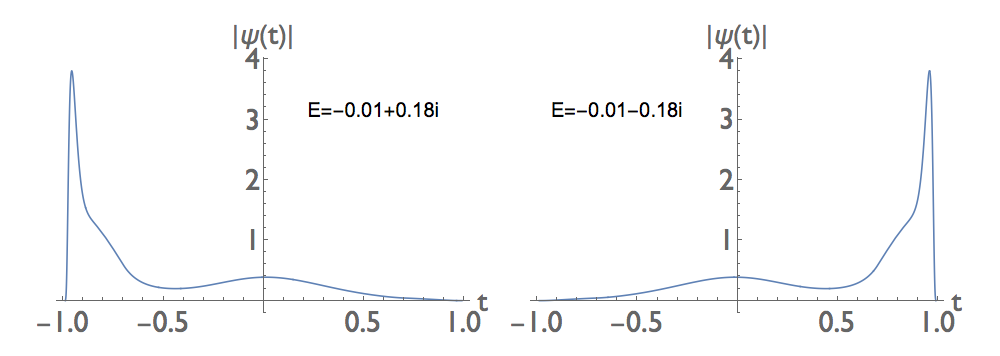}
\end{center}
\caption{[Color online] Absolute values of the eigenfunctions for the continuum
eigenvalues $-0.01\pm0.18i$ for $\vep=-2.6$. These eigenvalues belong to the
continuous spectrum. The indication that they are part of the continuous
spectrum is that at one of the boundary points the eigenfunctions suddenly drop
to $0$ rather than decaying exponentially to $0$. As in Fig. \ref{F13},
the left and right panels are interchanged under $t\to-t$, which corresponds
to a $\cPT$ reflection.}
\label{F14}
\end{figure}

\begin{figure}[t!]
\begin{center}
\includegraphics[scale=0.22]{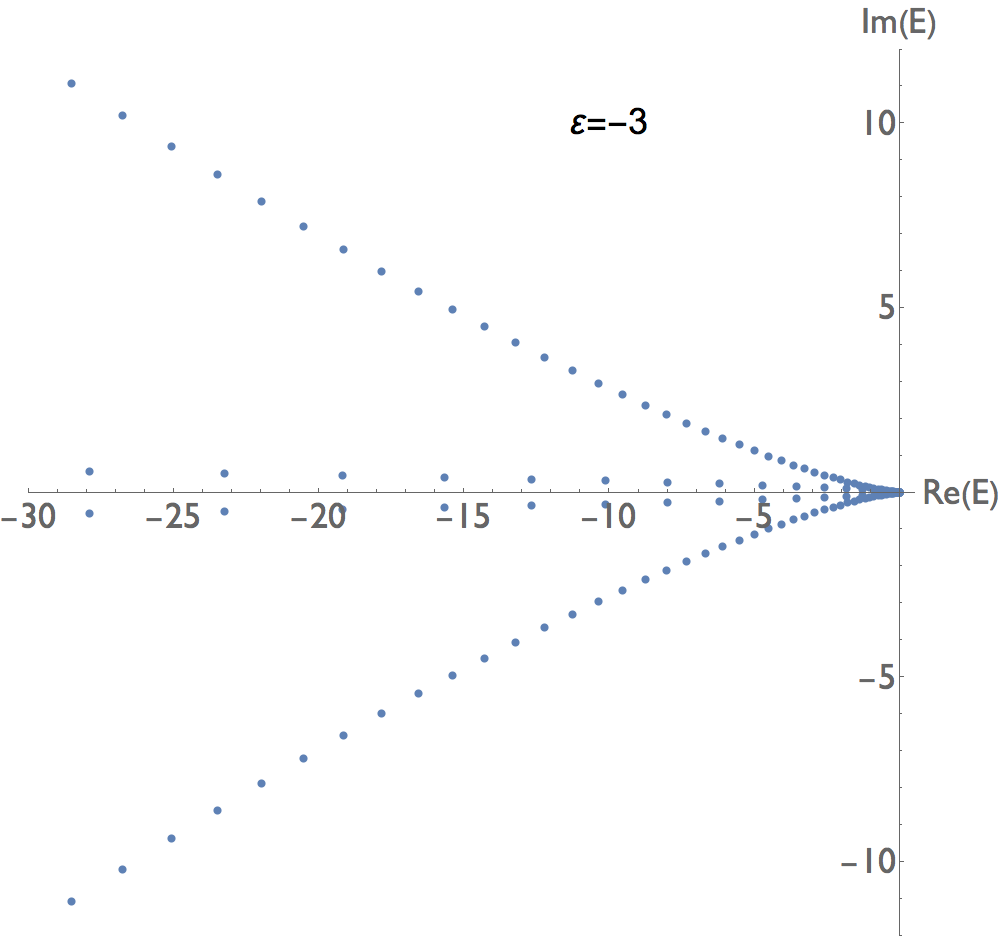}
\end{center}
\caption{[Color online] Eigenvalues for the Coulomb case $\vep=-3$.
There are no discrete eigenvalues and the continuum eigenvalues lie on four
curves in the left-half complex plane.}
\label{F15}
\end{figure}

\begin{figure}[h!]
\begin{center}
% \includegraphics[trim=26mm 15mm 0mm 12mm,clip=true,scale=0.31]{Fig16.pdf}
% trim=left bottom right top
\includegraphics[scale=0.22]{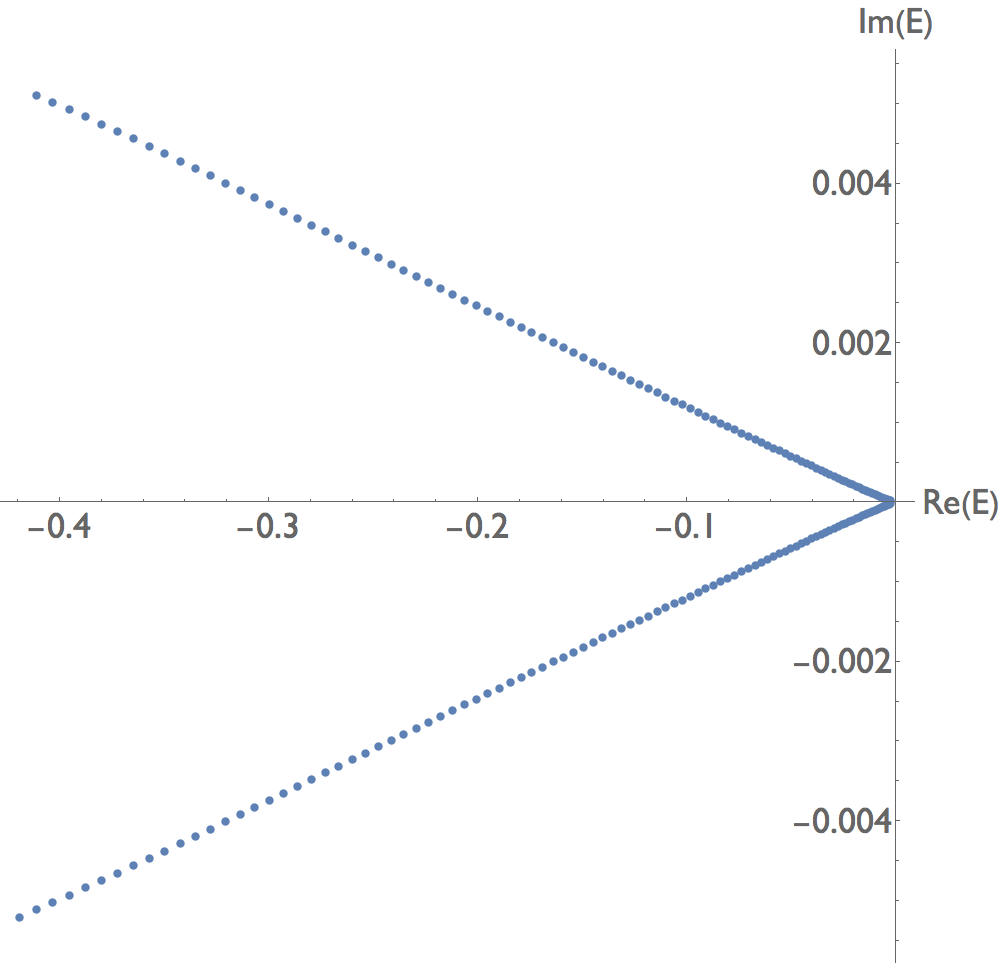}
\end{center}
\caption{[Color online] Detail of the region around the origin in the complex
eigenvalue plane of Fig.~\ref{F15} for $\vep=-3$. For this figure we have
chosen $\eta=0.999$ and have taken the very small cell size $0.00001$.}
\label{F16}
\end{figure}

For the Coulomb potential $\vep=-3$, (\ref{e30}) becomes
$$\psi''(s)-\frac{1}{s}\psi(s)=E\psi(s),$$
which is a special case of the Whittaker equation
$$w''(z)+\left[-\frac{1}{4}+\frac{\kappa}{z} + \frac{\fourth-\mu^2}{z^2}\right]
w(z)=0$$
with $\mu^2=\fourth$ \cite{R24}. The boundary conditions are unusual (they
differ from those in conventional atomic physics) in that $\psi(s)\to0$ as
$|s|\to\infty$ with ${\rm arg}(s)=\pm2\pi$. Rather than performing an
analytic solution to the eigenvalue problem, we simply present the numerical
results, which are obtained by solving (\ref{e32}) with $\vep=-3$. Figure
\ref{F15} displays about 100 eigenvalues, which lie on two pairs of
complex-conjugate curves in the left-half plane. These eigenvalues are part
of the continuous spectrum. A blow-up of the region
around the origin is shown in Fig.~\ref{F16}.

\begin{figure}[h!]
\begin{center}
\includegraphics[scale=0.22]{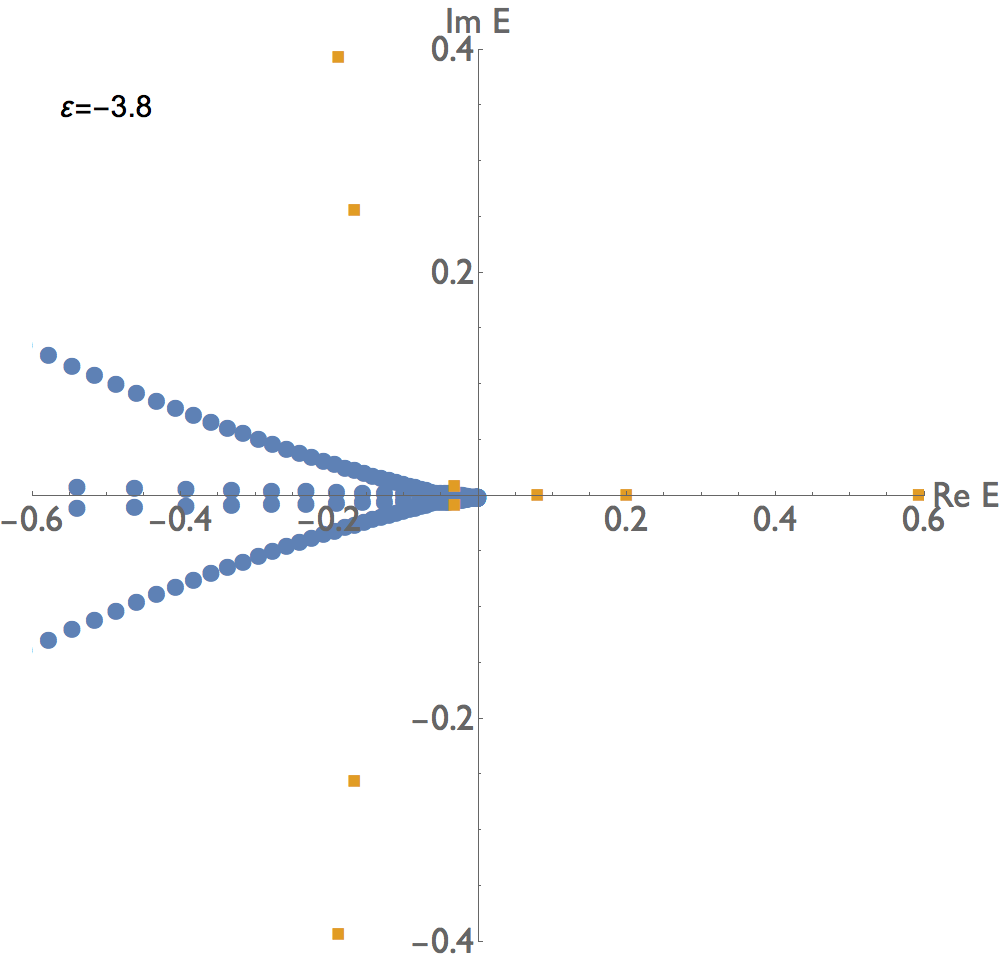}
\end{center}
\caption{[Color online] Eigenspectrum for $\vep=-3.8$. The continuous part
of the spectrum (blue dots) lies on two complex-conjugate pairs of curves in
the left-half plane and resembles that of the Coulomb case (see
Fig.~\ref{F15}). The discrete part of the spectrum (orange squares) consists of
complex-conjugate eigenvalues in the left-half plane and real eigenvalues on
the positive-real axis.}
\label{F17}
\end{figure}

\begin{figure}[h!]
\begin{center}
\includegraphics[scale=0.22]{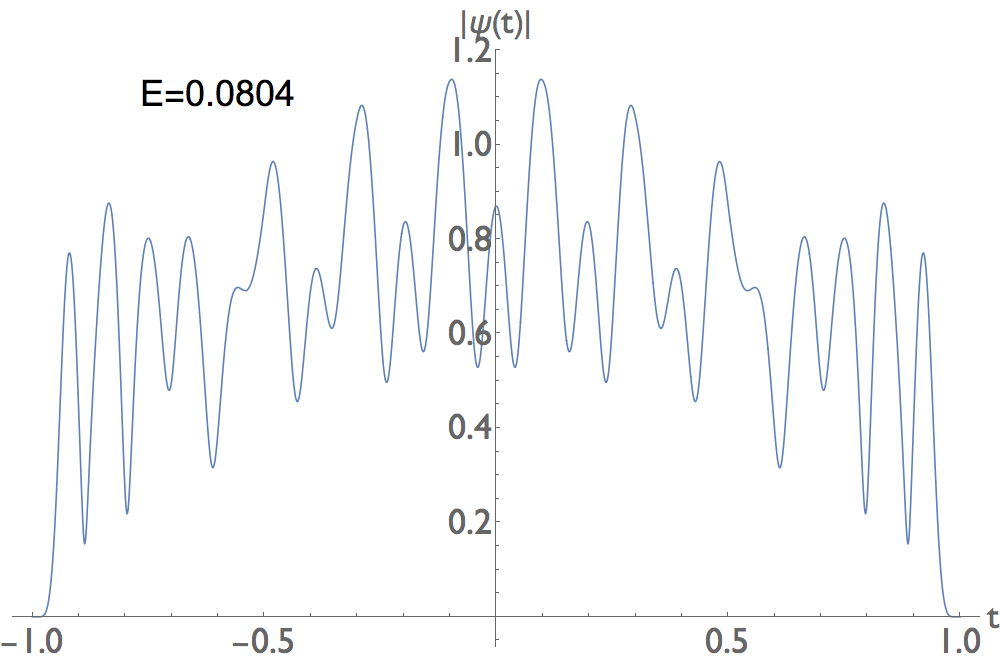}
\end{center}
\caption{[Color online] Plot of the absolute value of the eigenfunction
associated with the discrete real eigenvalue $E=0.0804$ for $\vep=-3.8$.}
\label{F18}
\end{figure}

The Coulomb case $\vep=-3$ is a transition point between the regions $\vep>-3$
and $\vep<-3$. In the first region the discrete eigenvalues occur in
complex-conjugate pairs and there are no real discrete eigenvalues (as we see in
Fig.~\ref{F11}). In the region $\vep<-3$ the discrete spectrum includes both
real and complex-conjugate pairs of eigenvalues in addition to the continuous
spectrum. Figure~\ref{F17} illustrates the typical distribution of eigenvalues
in the latter region for the choice $\vep=-3.8$. In Fig.~\ref{F18} we display
the eigenfunction for the real discrete eigenvalue $E=0.0804$. Unlike the
eigenfunctions in Figs.~\ref{F13} and \ref{F14} this eigenfunction is symmetric
in $t$. 

\subsection{Conformal limit $\vep\to-4$}

\begin{figure}[h!]
\begin{center}
\includegraphics[trim=60mm 48mm 60mm 50mm,clip=true,scale=0.64]{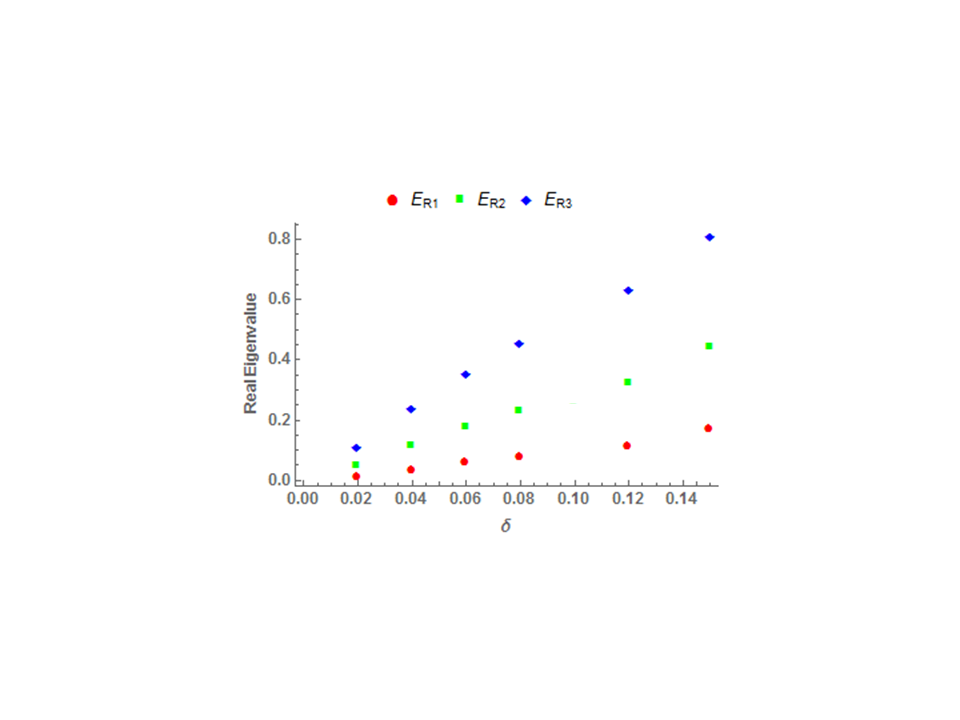}
\end{center}
\caption{[Color online] First three real eigenvalues of the Hamiltonian
$H=p^2+x^2(ix)^\vep$ plotted as functions of the parameter $\delta$, where
$\vep=-4+\delta$. The eigenvalues clearly approach $0$ as $\delta\to0$
and we see strong evidence that the eigenvalues vanish linearly with $\delta$.}
\label{F19}
\end{figure}

The limit $\vep\to-4$ is the conformal limit of the theory and thus the behavior
of the eigenvalues in this limit is interesting to determine. It is difficult to
study this limit because the eigenvalue equation in the complex-$s$ plane
follows a contour that loops around the origin many times when $\vep$ is near
$-4$. Indeed, the number of loops approaches $\infty$ as $\vep\to-4$ and, as a
consequence, we are less confident about the dependability of the Arnoldi
algorithm that we are using to obtain our numerical results. Nevertheless, we
have studied the spectrum for values of $\vep$ that are slightly greater than
$-4$ and examine the trend as $\vep$ moves closer to $-4$. We find that in this
limit the entire spectrum collapses to the origin. It is not easy to demonstrate
this by studying the continuous part of the spectrum; these points merely become
denser in the vicinity of the origin. However, the discrete eigenvalues move
toward the origin as $\vep\to-4$. In Table \ref{t3} we show the behavior of the
first three real eigenvalues as $\delta\to0$, where $\vep=-4+\delta$. These data
are plotted in Fig.~\ref{F19}. This figure suggests that the eigenvalues vanish
linearly with $\delta$.

\begin{table}[h!]
\begin{center}
\begin{tabular}{|c|c|c|c|}
\hline
$\delta$ & First real & Second real & Third real\\
         & eigenvalue & eigenvalue  & eigenvalue\\
\hline
0.15     & $0.173$ & $0.440$ & $0.807$ \\
0.12     & $0.114$ & $0.321$ & $0.628$ \\
0.08     & $0.080$ & $0.230$ & $0.454$ \\
0.06     & $0.060$ & $0.177$ & $0.351$ \\
0.04     & $0.035$ & $0.116$ & $0.236$ \\
0.02     & $0.012$ & $0.049$ & $0.106$ \\
\hline
\end{tabular}
\end{center}
\caption{\label{t3} First three real discrete eigenvalues as a function of
$\delta$, where $\vep=-4+\delta$. All the eigenvalues approach 0 as $\delta\to
0$. In fact, Fig.~\ref{F19} indicates that they approach zero in a linear
fashion.}
\end{table}

\section{Conclusions}
\label{s5}

In this paper we have studied the eigenvalues of $H$ in (\ref{e1}) for $-4<\vep<
0$ and we have shown that there is a rich analytic structure as a function of
the parameter $\vep$. We have identified transition points at the integer values
$\vep=0,\,-1,\,-2,\,-3$. Just above $\vep=0$ the eigenvalues are all
real and positive but below $\vep=0$ the eigenvalues split sequentially into
complex-conjugate pairs and all of the eigenvalues but one are complex below
about $\vep=-0.58$. At $\vep=-1$ the real parts of the eigenvalues approach
$\infty$ but the imaginary parts of the eigenvalues all vanish.

Below $\vep=-1$ the eigenvalues are once again finite, but as $\vep$ approaches 
$-2$ the entire spectrum coalesces to the value $-1$. Below $\vep=-2$ the
eigenvalues explode away from the value $-1$ and a new feature of the spectrum
arises: The spectrum is partly continuous and partly discrete. The continuous
part of the spectrum lies along complex-conjugate pairs of lines in the complex
plane that begin near the origin and run off to $\infty$. By contrast, the
eigenvalues belonging to the discrete part of the spectrum have negative real
parts.

At the Coulomb value $\vep=-3$ the continuous parts of the spectrum swing around
to the negative complex plane and the discrete eigenvalues disappear. Below the
Coulomb transition the discrete eigenvalues reappear and some of the discrete
eigenvalues are now {\it real}. As $\vep$ approaches the conformal point $-4$,
the spectrum appears to implode to the origin.

\acknowledgments
We thank S.~Sarkar for initial discussions of this investigation. CMB thanks the
Heidelberg Graduate School of Fundamental Physics for its hospitality and CS and
ZW thank the Physics Department at Washington University for its hospitality.


\begin{thebibliography}{99}
\bibitem{R26} The appearance of a singularity in the eigenspectrum when the
power of $x$ in the potential is $1$ is reminiscent of the singularity of the
Riemann zeta function $\zeta(z)$ at $z=1$.
\end{thebibliography}
\end{document}